\documentclass[twocolumn,aps,superscriptaddress,multicol,amsmath,amssymb]{revtex4-2}
\usepackage{amssymb}
\usepackage{amsmath}
\usepackage{graphicx}
\usepackage{epstopdf}
\usepackage{bm}
\usepackage{gensymb}
\usepackage{soul}
\usepackage{sidecap}
\usepackage[normalem]{ulem}
\usepackage{times}

\graphicspath{{Figures/}}
\usepackage{physics}
\usepackage{enumitem}
\usepackage[normalem]{ulem}

\newcommand{\be}{\begin{equation}}
\newcommand{\ee}{\end{equation}}
\newcommand{\bk}{{{\bf{k}}}}
\newcommand{\bq}{{{\bf{q}}}}

\newcommand{\bv}{{{\bf{v}}}}

\newcommand{\bea}{\begin{eqnarray}}
\newcommand{\eea}{\end{eqnarray}}

\newcommand{\bd}{\begin{displaymath}}
\newcommand{\ed}{\end{displaymath}}
\newcommand{\ba}{\begin{array}}
	\newcommand{\ea}{\end{array}}
\newcommand{\bi}{\begin{itemize}}
	\newcommand{\ei}{\end{itemize}}
\newcommand{\bc}{\begin{center}}
	\newcommand{\ec}{\end{center}}
\newcommand{\bfl}{\begin{flushleft}}
	\newcommand{\efl}{\end{flushleft}}
\newcommand{\bfr}{\begin{flushright}}
	\newcommand{\efr}{\end{flushright}}
\newcommand{\no}{\nonumber}

\newcommand{{\mi}}{\rm i}

\newcommand{\bl}{\begin{aligned}}
	\newcommand{\el}{\end{aligned}}

\def\bk{{\bf k}} \def\bq{{\bf q}}  
  \def\bd{{\bf d}}

\def\6{\partial}

\def\o{\omega}

\def\={\!\!\!&=&\!\!\!}
\def\+{\!\!\!&&\!\!\!+~}
\def\-{\!\!\!&&\!\!\!-~}

\graphicspath{{.}{./new_figs/}}

\usepackage{float}
\usepackage[caption = false]{subfig}
\usepackage{subfig}
\usepackage{enumerate}
\usepackage{multirow}
\usepackage{tabularx}
\usepackage{array}
\usepackage{url}
\usepackage{slantsc}
\usepackage{lmodern}
\newcommand\redout{\bgroup\markoverwith{\textcolor{red}{\rule[.5ex]{2pt}{0.4pt}}}\ULon}

\usepackage[colorlinks=true,citecolor=blue]{hyperref}
\hypersetup{colorlinks=true,citecolor=blue,linkcolor=red,urlcolor=blue}

\usepackage{hyperref,cleveref}
\begin{document}
\title{
 Polar Kerr Effect in Multiband Spin-Orbit Coupled Superconductors 
	}
\author{Meghdad Yazdani-Hamid}
\affiliation{Department of Physics, Bu-Ali Sina University, 65178, 016016, Hamedan, Iran}
%
\author{Mehdi Biderang}
\affiliation{Department of Physics, University of Toronto, 60 St. George Street, Toronto, Ontario, M5S 1A7, Canada}
\affiliation{Asia Pacific Center for Theoretical Physics (APCTP), Pohang, Gyeongbuk, 790-784, Korea}
\author{Alireza Akbari}
%
\affiliation{Asia Pacific Center for Theoretical Physics (APCTP), Pohang, Gyeongbuk, 790-784, Korea}
\affiliation{Max Planck POSTECH Center for Complex Phase Materials,and Department of Physics, POSTECH, Pohang, Gyeongbuk 790-784, Korea}
\affiliation{Institut für Theoretische Physik III, Ruhr-Universität Bochum, D-44801 Bochum, Germany
}
\date{\today}
%
\begin{abstract}
{
We undertake a  theoretical analysis 
to probe the Kerr spectrum within the superconducting phase of strontium ruthenate, where the
Kerr rotation experiments demonstrate the existence of a superconducting state with broken time reversal symmetry.
We find that  spin-orbit coupling  changes the hybridization along the Fermi surface's diagonal zone mainly affects Hall transport. We show that the dominant Hall response arises mainly from the quasi-1D orbitals $d_{yz}$ and $d_{xz}$, linked to their hybridization, while other contributions are negligible. 
This establishes that,  the breaking of time reversal symmetry of quasi-1D orbitals can account for the existence of the Kerr angle, irrespective of  the order symmetry specific to the $d_{xy}$ orbital.
Moreover,
the optical Hall conductivity and Kerr angle estimated for the hypothesised superconducting orders also closely match the experimental findings, providing important novel insight on the role of the spin-orbit coupling, hybridization, and emergent order.
}
\end{abstract}
%
\maketitle
%

%

\section{Introduction}
\label{Sec:I}
Despite extensive efforts and a multitude of precise experiments, the true nature of the superconducting order parameter in Sr$_2$RuO$_4$ continues to elude a definitive answer~\cite{benhabib2021ultrasound,chronister2021evidence,ghosh2021thermodynamic,
grinenko2021split,grinenko2021state,kivelson2020proposal,romer2021superconducting,Bhattacharyya2023Superconducting,Kaser2022Interorbital,Ikegaya2020Identification,Gingras2022Superconductivity,Wagner2021Microscopic,Lindquist2020Distinct,Lindquist2023Reconciling,Willa2021Inhomogeneous,Romer2022Leading}.
From an alternative perspective, there is no single comprehensive universal order parameter  that can explain all of those outcomes of conducted experiments.
For more than two decades,
its superconductivity state  has been viewed as a two-dimensional analogy to the
super-fluid state of $^3$He, namely two-dimensional $p_x$+${\mi}p_y$ (a time-reversal
symmetry-breaking superconducting state due to spontaneous magnetic fields)~\cite{rice1995sr2ruo4}.
The initial report of the nuclear magnetic resonance (NMR)  measurement of the Knight shift~\cite{luke1998time}
and the existence of a non-zero Kerr angle~\cite{xia2006high} are known as two strong experimental evidences that support this
proposed superconducting pairing. However, the lack of edge currents~\cite{scaffidi2015large} and the absence of a cusp in the strain
dependency of $T_c$~\cite{hicks2014strong,steppke2017strong} exclude this odd-parity pairing state. The more recent NMR experiments have been shown to be inconsistent with triplet
pairings and have revealed that the dominant pairings have spin-singlet character and suggest time-reversal symmetry breaking
linear combinations of  extended s-wave~\cite{ishida2020reduction}.

One of the main reasons for the aforementioned longstanding issue is the multi-orbital nature of Sr$_2$RuO$_4$ that hampers the identification of active and passive bands for the superconducting pairing. The quasi-2D nature of the electronic dispersion introduces these two classes and represents that the superconducting order is derived from either the quasi-2D orbital $d_{xy}$ or the quasi-1D orbitals $d_{xz}$ and $d_{yz}$~\cite{agterberg1997orbital}. Two classes possess fully different Fermi surfaces such that for the former case, it is isotropic (one circular Fermi surface) whereas for the latter the resulting Fermi surface is highly anisotropic due to near degeneracies and avoided crossings of different bands. Some thermodynamic and transport properties may favor the first class as the active band and others the second one. For example,
the model used
under biaxial or uniaxial pressure is based on the first band class because Lifshitz transition occurs for the quasi-2D orbital d$_{xy}$ whose Van Hove peak is nearest to the Fermi surface and have an isotropic Fermi sheet~\cite{Cobo2016,barber2018resistivity,Yu2020Critical,Kim2023Suppressed}. The anisotropic Fermi surfaces of the quasi-1D orbitals $d_{xz}$ and $d_{yz}$ with the hybridization gives rise to a finite Berry curvature which have a close connection with Hall-type response~\cite{robbins2017effect,Gradhand_2014}. In fact,  Berry curvature is associated with the orbital hybridization. Taking into account SOC provides a spin dependent hybridization between three orbitals and gives rise to mixing two classes. Thus one can expect that SOC has profound effects on the Hall-type responses~\cite{robbins2017effect}.

There are additional supporting evidences indicating the predominant involvement of quasi-1D orbitals, namely 
from inelastic neutron scattering studies~\cite{Sidis1999Evidence}, exact weak-coupling analysis of the Hubbard model~\cite{raghu2010hidden}, and tunneling spectroscopy results\cite{firmo2013Evidence}. Nevertheless, it is noteworthy that these orbitals seem to exert minimal influence on heat capacity measurements~\cite{Deguchi2004Gap} and remain relatively unaffected when subjected to strain~\cite{barber2018resistivity}. This can be attributed to the proximity of the $d_{xy}$ orbital to the van Hove singularity. Consequently, it becomes increasingly evident that the quasi-1D orbitals play a pivotal role, at the very least, in governing the intrinsic anomalous Hall transport. Identifying these one-dimensional orbitals as the primary active entities also elucidates the absence of edge currents~\cite{Bjornsson2005Scanning, Kirtley2007Upper, Hicks2010Limits}.

In the present work, we theoretically study intrinsic Hall response of the superconducting state of Sr$_2$RuO$_4$, using the three-orbital model in the presence of SOC.
So far, this signal has been investigated using  particular  approaches, including single-orbital models with impurity scattering~\cite{goryo2008impurity, kim2008hall, lutchyn2009frequency, konig2017kerr}, two-orbital models without impurities~\cite{taylor2012intrinsic, wysokinski2012intrinsic,denys2021origin,ifmmode_2012,Zhang2023Quantum,Liu2023Impact}, as well as first-principle calculations~\cite{gradhand2013kerr, robbins2017effect}.
Extending the procedure outlined in Ref.~\cite{taylor2012intrinsic}, we have determined that the primary contribution to the optical Hall conductivity, and consequently, the polar Kerr effect, emanates from the quasi-1D orbitals, even when considering interorbital interactions within a mixture of two classes and accounting for the rearrangement of orbital hybridization.
In other words, at least one of the quasi-1D orbitals needs to have its superconducting time reversal symmetry broken in order to achieve a suitable Kerr rotation.

In the following section, we outline the physical model used to describe the electronic band structure and superconducting ground state. Section \ref{PKE} will investigate the dynamical Hall conductivity and calculate the polar Kerr angle based on the most recent suggested superconducting gap functions for 
 Sr$_{2}$RuO$_{4}$.
Finally, the last section provides a short summary and conclusion.
%

%
\section{Model Hamiltonian}
\label{Sec:Model}
Employing the effective three-orbital model that encompasses the t$_{2g}$ orbital manifold, 
the Hamiltonian for normal state can be written as
follows~\cite{Ramires2016Identifying,Ramires2017Notes,Kallin2020Spin}
\be
\bl
&
{\cal H}_0=
\sum_{\bk }
\psi^{\dagger}_{\bk }
{ \hat{\cal  H}}(\bk)
\psi^{}_{\bk  },
\el
\ee
%
where
  ${ \hat  {\cal  H}}(\bk)$ is defined as
%
\begin{align}
 { \hat  {\cal  H}}(\bk
)=
\begin{pmatrix}
\xi_{\bk}^{yz} \sigma_{0} 
&
{\rm g}_{\bk}\sigma_{0}
+{{\mi}} \lambda \sigma_{z}
&
-{{\mi}} \lambda\sigma_{y}
\\
{\rm g}_{\bk} \sigma_{0} 
-{{\mi}} \lambda\sigma_{z}
&
\xi_{\bk}^{xz} \sigma_{0} 
&
{\mi} \lambda \sigma_{x}
\\
{{\mi}} \lambda\sigma_{y}
&
-{{\mi}} \lambda\sigma_{x}
&
\xi_{\bk}^{xy}\sigma_{0} 
\end{pmatrix},
\label{Eq:Ham}
\end{align}
under the basis
$\psi^{\dagger}_{\bk }=(\phi^\star_{\bk\uparrow},\phi^\star_{\bk\downarrow})$,
with
$\phi^\star_{\bk\sigma}=(d^{\dagger}_{yz,\bk\sigma},d^{\dagger}_{xz,\bk\sigma},d^{\dagger}_{xy,\bk{\bar \sigma}})$,
and
 $\sigma_{i=x,y,z}$ denotes the $2 \times 2$ Pauli matrices in the spin basis.
The field operator $d^{\dagger}_{ \nu,\bk\sigma}$ ($d^{}_{ \nu,\bk\sigma}$) creates (annihilates) an electron with momentum $\bk$ and spin $\sigma$ (${\bar \sigma}=-\sigma$) at orbital $\nu=yz,xz,xy$.
Moreover, $\lambda$ is the strength of SOC
 and the electronic dispersions are described by the following tight-binding relations~\cite{akbari2013}
%
\begin{align}
\begin{aligned}
\no
\xi_{\bk}^{yz}
&
=
-\mu_1
-2t^{yz}_{x}
\cos k_x
-2t^{yz}_{y}
\cos k_y,
\\
\xi_{\bk}^{xz}
&
=
-\mu_1
-2t^{xz}_{x}
\cos k_x
-2t^{xz}_{y}
\cos k_y,
\\
\xi_{\bk}^{xy}
&
=
-\mu_2
-2t^{xy}_{x}
\cos k_x
-2t^{xy}_{y}
\cos k_y
-4t^{xy}_{xy}
\cos k_x
\cos k_y,
\\
{\rm g}_{\bk}
&
=
-2g
\sin k_x
\sin k_y,
\label{Eq:trigonometric_functions}
\end{aligned}
\end{align}
%
where $\mu_{i=1,2}$,  $t_{\vartheta=x,y,z,xy}^\nu$ and ${g}$ represent
 the chemical potentials,
hopping parameters and the hybridization coefficient, respectively.
Moreover,
the inter-orbital coupling $g$ and hopping integral $t^{xy}_{xy}$ correspond to the next-nearest neighbours. The tight-binding parameters are set as
$t_x^{xz}=t_y^{yz}=t=0.4$eV,
$t_y^{xz}=t_x^{yz}=0.1t$,
$t_x^{xy}=t_y^{xy}=0.8t$,
$t_{xy}^{xy}=0.3t$,
$g=0.1t$,
$\mu_1=t$,
$\mu_2=1.1t$~\cite{Cobo2016}.
%
%
%
%
Within the Nambu space $\psi'^{\dagger}_{\bk }\!\!=\!\!(\phi^\star_{\bk\uparrow},\phi^\star_{\bk\downarrow},\phi_{-\bk\uparrow},\phi_{-\bk\downarrow})$,  the superconducting Hamiltonian 
can be written as
%
\be
 {\cal  H}_{\rm SC}=\sum_{\bk} \psi'^{\dagger}_{\bk }{\hat {\cal H}}_{\rm SC}(\bk)\psi'^{}_{\bk },
\ee
%
where  ${\hat {\cal H}}_{\rm SC}(\bk)$
is a
$12\!\times\!12$ tensor defined by
%
\begin{align}
{\hat {\cal H}}_{\rm SC}(\bk)=
\begin{bmatrix}
{\hat  {\cal  H}}(\bk)
&\quad
{\hat \Delta}_{}
(\bk)
\\
{\hat \Delta}^{\dagger}
(\bk)
& \quad
-{\hat  {\cal  H}}^{*}(-\bk)\
\end{bmatrix},
\end{align}
%
in which the superconducting order parameter
 is a $6\times 6$
antidiagonal matrix
$
{\hat \Delta}(\bk)
\!=\!
{ \Delta}(\bk)
\otimes \sigma_x
$.
Here, ${\Delta}(\bk)$ is a $3\times 3$ matrix, and we exclusively focus on intra-orbital pairing within a diagonal matrix format, which comprises the following orbital elements
$\Delta^\nu_{\mathbf{k}}=\Delta^{\nu\prime}_{\mathbf{k}}+{\mi}\Delta^{\nu\prime\prime}_{\mathbf{k}}$.
Note that
the multi-orbital (band) structure makes different superconducting gap
 textures
  on each orbital (band).
Later,  we set the most favored forms for the orbital order parameters to calculate intrinsic anomalous Hall transport.

\section{Optical Hall conductivity and Polar Kerr Effect}
\label{PKE}
Within the linear response theory,
the dynamical Hall conductivity is given by the following Kubo formula
%
\begin{align}
\label{hh}
\begin{split}
\sigma^H(\omega)&=\frac{1}{2}  \lim_{\bq \to 0}[\sigma^{}_{xy}(\bq,\omega)-\sigma^{}_{yx}(\bq,\omega)],
\end{split}
\end{align}
with
\be
\sigma^{}_{ij}(\bq,\omega)=\frac{{\mi}}{\omega}K^{}_{ij}(\bq,\omega).
\ee
%
Here $K^{}_{ij}(\bq,\omega)$ is the current correlator and  is obtained from the analytical continuation, ${{\mi}}\omega^{}_m\!\rightarrow\! \omega+{{\mi}}0^+$, of its Matsubara counterpart
%
\be
\bl
\label{correlation}
&K^{}_{ij}(\bq,{{\mi}}\omega_m)
=
\\
&\quad\;
T \sum_{\bk,{{\mi}} {\upsilon}_n}
{\rm Tr}
\Big[
{\hat {\cal J}}_i(\bk)
{\hat {\cal G}}_0(\bk,i {\upsilon} _n)
{\hat {\cal J}}_j(\bk)
{\hat {\cal G}}_0(\bk+\bq,
{{\mi}}  {\upsilon}_n
\!+\!
{{\mi}}{\omega}_m)
\Big],
\el
\ee
%
in which ${\upsilon}_n\!\!=\!2(n+1)\pi T$, and ${\omega}_m\!=2m\pi T$ are Matsubara fermionic and bosonic frequencies, respectively.
In the above equation, the bare Green's function in the superconducting
state is a 12$\times$12 matrix (2 Nambu, 3 orbital,
2 spin degrees of freedom) and is given by
\be
\bl
\label{G-func}
{\hat {\cal G}}_0(\bk,{{\mi} }{\omega} _m)&=
\Big[
{{\mi} }{\omega} _m-{\hat {\cal H}}_{\rm SC}(\bk)
\Big]^{-1}
\\
&=\mqty*[\hat{G}_0(\bk,{{\mi}}
\omega_m)
&
\hat{F}_0(\bk,{{\mi}}\omega_m)
\\
\hat{F}^\dag_0(\bk,{{\mi}}\omega_m)
&
-\hat{G}^{\intercal}_0(-\bk,-{{\mi}}\omega_m)],
\el
\ee
%
%
in which
$\hat{G}_0(\bk,{{\mi}}\omega_m)$
and
$\hat{F}_0(\bk,{{\mi}}\omega_m)$
are the normal and anomalous Green's functions, respectively.
%
%
In Eq.~(\ref{correlation}), the charge current operator ${\hat{\cal J}_i}(\bk)$ is defined as
\be
\bl
{\hat{\cal J}_i}(\bk)
=
e \hat{ \mathbf{v}}_i (\bk)
={\hat J}^{(e)}_i(\bk) \oplus {\hat J}^{(h)}_i(\bk),
\el
\ee
%
wherein
\be
\bl\no
{\hat J}^{(e)}_i(\bk)=-e \nabla_{k_i}    { \hat  {\cal  H}}(\bk)   ;
\;\;
\;\;
{\hat J}^{(h)}_i(\bk)= e \nabla_{k_i}    (-{\hat {\cal H}}^{*}(-\bk))
\el
\ee
are particle and hole contributions of the charge current operator, respectively.
Taking into account the bare current vertex tensor $\hat{ \mathbf{v}}_i(\bk)$,
and substituting the above terms into
Eq. (\ref{hh}), 
we can show that the Hall conductivity is governed  by the following proportionality equation
\be
\sigma^H(\omega)^{}
\propto
(\bv^{xz}-\bv^{yz})\times \bv^{xz-yz},
\ee
with
$$\bv^{xz(yz)}=
\partial_{k_x}\xi_{\bk}^{xz(yz)}\hat{i}
+\partial_{k_y}\xi_{\bk}^{xz(yz)}\hat{j};
$$
and
$$\bv^{xz-yz}=
\partial_{k_x}{\rm g}_{\bk}\hat{i}
+\partial_{k_y}{\rm g}_{\bk}\hat{j},
$$
which  will end up with different  set of
the prefactor components of the velocity.
Terms including $\bv^{xy}$, are proportional to
$\lambda^2  g $
that in comparison with $\bv^{yz}$ and $\bv^{xz}$ with dependency as $g$, are negligible. Therefore, the dominant terms is related to the prefactors of $v_{x}^{yz}v_{y}^{xz-yz}$, $v_{y}^{yz}v_{x}^{xz-yz}$, $v_{x}^{xz}v_{y}^{xz-yz}$ and $v_{x}^{yz}v_{x}^{xz-yz}$
(see Appendix \ref{coefficient}). In fact, these results express that the leading contribution to the Hall-type response originates from the $d_{yz}$ and $d_{xz}$  orbitals even in the presence of SOC which couples the orbital $d_{xy}$ with these orbitals.
Defining 
$z={{\mi}}\omega_m$ and 
$z^\prime={\mi}(\omega_m+\upsilon_n)$ 
and in accordance with the explanations provided earlier, the dynamical Hall conductivity can be expressed as follows
\be
\bl
\label{hall:conduc}
\sigma^H(\omega)
&=
\frac{2\mi e^2}{\omega}
\sum_{\bk}
{\rm g}_{\bk}
\Im
[
\Delta^{xz*}_{\bk}\Delta^{yz}_{\bk}
]
\\
&
\quad
\times
\big[
(v_{x}^{xz}-v_{x}^{yz})v_{y}^{xz-yz}-(v_{y}^{xz}-v_{y}^{yz})v_{x}^{xz-yz}
\big]
\\
&
\quad
\times
\lim_{{{\mi}}\omega^{}_m\!\rightarrow \omega+{{\mi}}0^+}
\Bigg[
T\sum_{ {\mi}\upsilon_n}
(z^2-{z^\prime}^2)(z+{z^\prime})
\\
&\quad
\times
\frac{
(z^2-E^{yz 2}_{\bk})
(z^2-E^{xz 2}_{\bk})
(z^2-E^{xy 2}_{\bk})^2
}
{
(z^2-E^{ 2}_{1\bk})^2
(z^2-E^{ 2}_{2\bk})^2
(z^2-E^{ 2}_{3\bk})^2
}
\\
&\quad
\times
\frac{
({z^\prime}^2-E^{yz 2}_{\bk})
({z^\prime}^2-E^{xz 2}_{\bk})
({z^\prime}^2-E^{xy 2}_{\bk})^2
}
{
({z^\prime}^2-E^{ 2}_{1\bk})^2
({z^\prime}^2-E^{ 2}_{2\bk})^2
({z^\prime}^2-E^{ 2}_{3\bk})^2
}
\Bigg]
.
\el
\ee
%
%
%
This expression elucidates the conditions under which the superconducting state 
may display a nonvanishing Kerr rotation.
It implies that there must be a phase difference between the one-dimensional orbitals.
Moreover the superconducting state of these two orbitals, in accordance with the two-dimensional feature, must also break the time-reversal symmetry.
Additionally, a significant contribution comes from the essential role played by interorbital velocity, which is required for the existence of a non-zero anomalous Hall effect.
As a consequence, the quasi-1D orbitals with a $\bk$-dependent hybridization leads to a nonzero Hall transport.
Here we define
$E_{\bk }^{\nu}=\sqrt{\xi^{\nu 2}_{\bk}+|\Delta^{\nu}_{\bk}|^2}$ for the orbital $\nu$, 
and
the BCS quasiparticle spectra,
$E^{ }_{i\bk}$ ($i=\!1\!-3$),
can be obtained by rephrasing  $\mathbf{D}(\bk, z={{\mi} }{\omega} _m)=\det[\hat{G}^{-1}_0(\bk,{{\mi}}\omega_m)]$,~as
\be\bl
\mathbf{D}(\bk,z)
&
%
\propto
%
(z^2-E^{ 2}_{1\bk})^2
(z^2-E^{ 2}_{2\bk})^2
(z^2-E^{ 2}_{3\bk})^2
.
\el\ee
The minimum of the sum of the quasiparticle spectra occurs at the diagonal zone, where $k_x = k_y = \pi/2$.
This proximity to the diagonal zone is closely related to orbital hybridization, which plays a pivotal role in determining the frequency dependence of the dynamical Hall-type response in multi-orbital superconductors exhibiting broken time reversal symmetry.
Moreover, the resonance peak that appears at the minimum of the sum of the quasiparticle spectra is affected by the  intensity of SOC.
The presence of SOC  has a notable impact on orbital hybridization, particularly at the point $(\pi,\pi)$ where the Fermi surface intersects orbital sheets. Consequently, one can anticipate that this interaction affects the dynamical optical response.
However, in the absence of SOC, this minimum shifts to
$$
{\rm min} [
 E_{1\bk}+E_{3\bk}
]
\approx 2 g 
;
\quad
E_{2\bk}=0,
$$
aligning with the findings of Ref.~\cite{taylor2013anomalous}.
\\

The Hall conductivity behaviour and other electronic characteristics in the superconducting state are determined by the precise shape of the superconducting gap symmetry.
Therefore, to calculate $\sigma^H(\omega)$, we need to know the superconducting order parameter.
In the case of a two-dimensional square lattice material, the classification of gap symmetries is determined by the irreducible representations (irrep.) of the D$_{4h}$ group, which encompass 
extended s-wave, 
$g_{xy(x^2-y^2)}$,
$d_{x^2-y^2}$,
$d_{xy}$,  
and $p_{x(y)}$, as elaborated
\be
\bl
A_{1g}:&\quad
s^\prime=\cos k_x+\cos k_y,
\\
A_{2g}:&\quad
g_{xy(x^2-y^2)}=\sin k_x \sin k_y [\cos k_x -\cos k_y ],
\\
B_{1g}:&\quad
d_{x^2-y^2}=\cos k_x -\cos k_y ,
\\
B_{2g}:&\quad d_{xy}=\sin k_x \sin k_y ,
\\
E_u:&\quad p_{x(y)}=\sin k_{x(y)} .
\el
\ee
In this context, one noteworthy aspect is the capacity to elucidate signs of time reversal symmetry breaking, prompting our exploration of the following scenarios for superconducting gap momentum dependency~\cite{taylor2012intrinsic,scaffidi2020degeneracy}, defined as
$(\Delta^{yz}_{\bk},\Delta^{xz}_{\bk})$:
\be
\bl
&(i) \quad \;
\Delta_0
({\mi} \sin k_y \cos k_x, \sin k_x \cos k_y)
\\
&(ii) \quad
\Delta_0
( \cos k_x - {\mi} \cos k_x , \cos k_y + {\mi} \cos k_y)
\\
&(iii) \quad \!\!
\Delta_0
(\cos k_x+{\mi} \sin k_y \cos k_x ,\cos k_y+{\mi} \sin k_x \cos k_y),
\el
\ee
for $p+{\mi} p$, $s^\prime+{\mi} d$,
and 
$s^\prime+{\mi} p$, respectively.
Moreover, we set the pairing on  the d$_{xy}$ orbital as~\cite{kivelson2020proposal,Palle2023Constraints,Wang2022Higher}
\be
\Delta^{xy}_{\bk}=
\Delta^\prime_0(\cos k_x-\cos k_y)(1+{\mi} \sin k_x\sin k_y).
\ee
%

\begin{figure}[t]
 \center
\includegraphics[width= 1\linewidth]{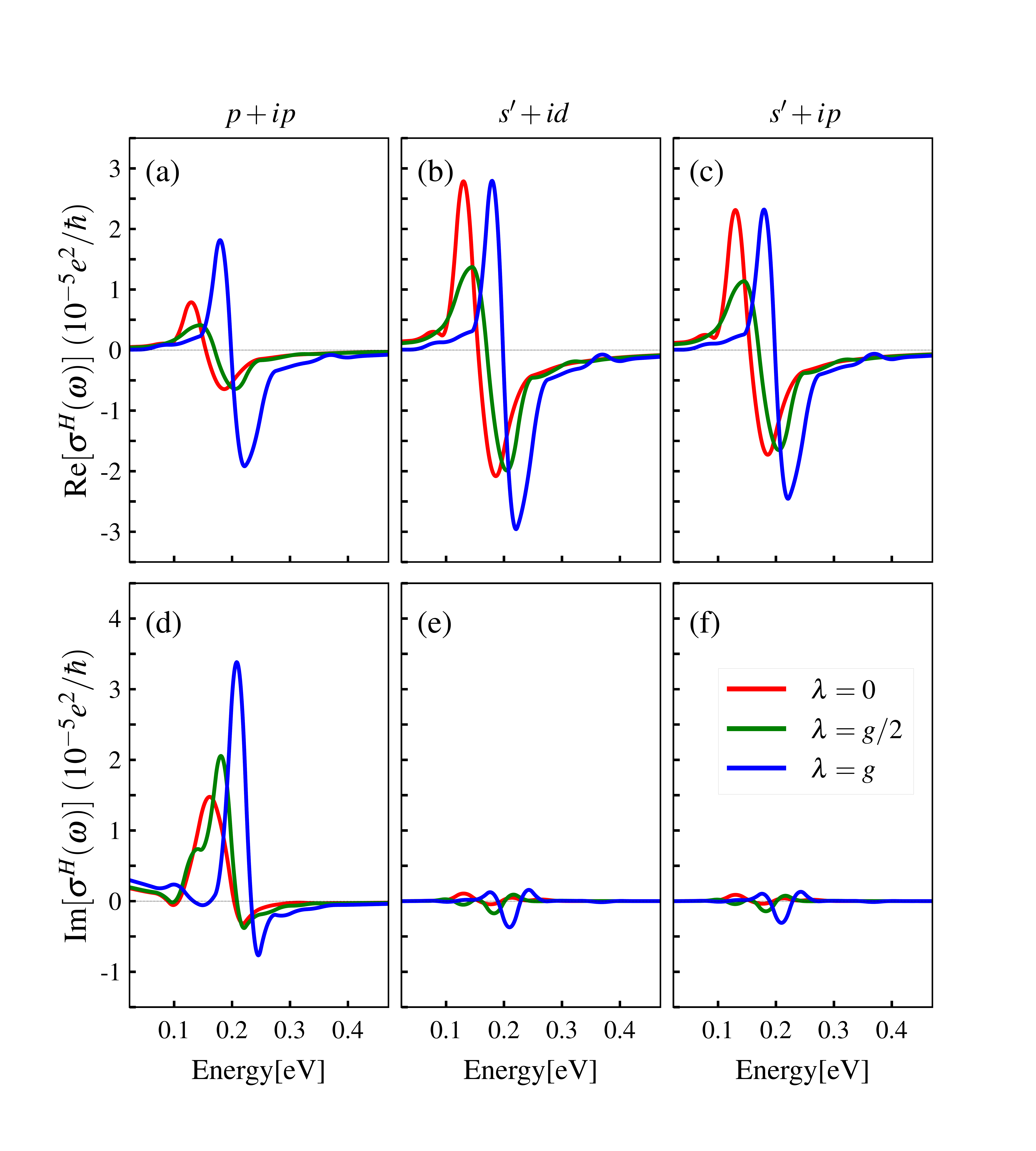}
\caption{
(a-c) Plotting the real parts and (d-f) imaginary parts of the dynamical Hall conductivity $\sigma^H(\omega)$
versus photon energy
at zero temperature for various quasi-1D pairings: $p+{\mi}p$,
$s^\prime+{\mi}d$, and 
$s^\prime+{\mi}p$,
with the $d+{\mi}g$ pairing for quasi-2D orbital, considering values of the spin-orbit coupling: $\lambda=0$ (red), $\lambda= g/2$ (green), and $\lambda= g$ (blue). 
}
\label{fig2}
\end{figure}
%
%

In Fig.~\ref{fig2},
we illustrate the results for the dynamical Hall conductivity at zero temperature, $\sigma^H(\omega)$, across three different strengths  of SOC.
The finite values of SOC are chosen as $\lambda= g/2$~\cite{yanase2003microscopic} and $\lambda= g$~\cite{haverkort2008strong,raghu2010hidden},
and we  set $\Delta^\prime_0=2\Delta_0=0.5$ meV~\cite{taylor2012intrinsic}.
In the lower panel, we observe $\Im\sigma^H(\omega)$, which characterizes the absorption spectrum. In the absence of  SOC for all pairings, the initiation of absorption processes, arising from orbital hopping between $d_{yz}$ and $d_{xz}$ orbitals, manifests at the interorbital interaction and peaks at the minimum point in the sum of the two lowest BCS quasiparticle spectra in the diagonal region,  i.e. 2$g$. 
Incorporating SOC results in a notable shift of both the onset and resonance peak toward higher frequencies. This shift may be related to the reconfiguration of hybridization between the quasi-1D orbitals, the introduction of transitions involving the quasi-2D orbital with the quasi-1D orbitals, and the energy splitting in regions characterized by degenerate $t_{2g}$ bands.
In fact, the frequency region
$2\Delta_0\lesssim\omega\lesssim2g+\lambda$
corresponds to interorbital hoppings, which can arise from either interorbital interactions or SOC.
The main peaks in this region are attributed to near degeneracies (along the diagonal zone) between orbitals in the quasiparticle band structure and the orbital mixing.
The positive feature seen at frequencies $\omega\lesssim0.1$ ($\omega\lesssim0.15$), as well as the negative features around $0.1$eV and $0.2$eV ($0.15$eV and $0.25$eV) for $\lambda=0$ ($\lambda\approx g$), can be assigned to the appearance of the $d_{xy}$  orbital and its hopping with the quasi-1D orbitals by comparing Fig.~\ref{fig2}(d-f) with the findings in Ref.~\cite{taylor2012intrinsic}.
Additionally, it is worth noting that the various pairings exhibit distinct resonance peak intensities and spectral weights, likely associated to their unique pairing characteristics. In the case of nonchiral pairings, the absorption spectrum tends to be smaller compared to the chiral case. A substantial reduction in spectral weight occurs at higher frequencies, approximately $\omega\gtrsim0.4$eV, which corresponds to associated with strongly correlated electronic limit.

%
%

Utilizing the dynamical Hall conductivity and following the established formalism of Kerr rotation one can obtain the polar Kerr angle as
%
\be
\bl
\theta^{\rm Kerr}(\omega)=
\frac{4\pi }{\omega d}
\Im
\Big[
\sigma^{H}(\omega)\varphi(\omega)
\Big],
\el
\ee
 %
where $d=6.8 {\AA}$ is the interlayer distance~\cite{lutchyn2009frequency}, and
%
$$\varphi(\omega)=
\Big[
n(\omega)
[n(\omega)^2-1]
\Big]^{-1}.
$$
%
Here $n(\omega)=\sqrt{\epsilon_\infty+(4\pi i/\omega)\sigma(\omega)}$ denotes the refraction index   defined by the dynamical Drude model,
%
$\sigma(\omega)=\sigma_0/(1-i\omega\tau)$,
%
where $\sigma_0$ 
 is the static Drude conductivity  with relaxation time
  $\tau^{-1}=\gamma_{\rm^{}_{Scatt}}=0.4$eV,
  and
  $\epsilon_\infty =10$ refers to the background dielectric tensor.
 Notice that Kerr angle strongly depends on
the quasiparticle scattering rate, $\gamma^{}_{\rm _{Scatt}}$~\cite{goryo2008impurity,lutchyn2009frequency}.

\begin{figure}[t]
 \center
\includegraphics[width= \linewidth]{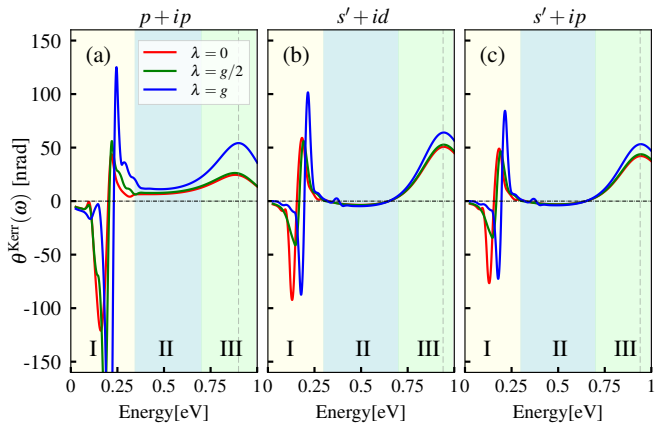}
\caption{
Plotting the Kerr rotation angle (in unit of nanoradian) versus photon energy for three pairing symmetries: $p+{\mi} p$, $s^\prime+{\mi} d$, and $s^\prime+{\mi} p$, for three values of spin-orbit coupling: $\lambda=0$, $\lambda= g/2$, and $\lambda= g$. 
We can roughly categorize the behavior of the Kerr rotation angle into three distinct regimes: I, II, and III.
The region I represents interorbital transitions, and the  vertical dashed line indicates the maximum of angle in the region III. 
}
\label{fig3}
\end{figure}
%

%
%
The results of Kerr rotation angle versus the photon energy  are shown in
Fig.~\ref{fig3},  covering  different strengths of SOC and for pairing symmetries: (a) $p+{\mi} p$, (b) $s^\prime+{\mi} d$, and (c) $s^\prime+{\mi} p$.
In a qualitative sense, we can delineate three primary regions labeled as I, II, and III, each represented by distinct colors.
The region I, encompassing frequencies $\omega\lesssim0.3$eV, reveals a complex structure with the resonant enhancements which are associated with interorbital transitions.
In the intermediate frequency range (region II), the Kerr rotation angle exhibits a distinctive flat-like behavior.
The variation in this region can be attributed to the fact that the primary contribution to $\theta^{\rm Kerr}(\omega)$ for nonchiral pairings arises from the real part of $\sigma^H(\omega)$ in the strongly correlated system limit. In contrast, for the chiral pairing $p+{\mi} p$, both the real and imaginary parts of $\sigma^H(\omega)$ are of similar magnitude, resulting in comparable contributions to $\theta^{\rm Kerr}(\omega)$.
On the other hand, region III shows a hump that corresponds to the situation where the spectra are dominated by a strong magneto-optical response at $\omega<\omega_{\rm edge}$ for chiral pairing and at $\omega>\omega_{\rm edge}$ for nonchiral pairings, respectively.
Here 
$\omega_{\rm edge}=\omega_{p}/\sqrt{\epsilon_\infty}$ 
is the frequency of the plasma edge,
and $\omega_p$ is the plasma frequency 
 and we set it as $\omega_p=2.9$eV~\cite{taylor2012intrinsic, lutchyn2009frequency}.
 The behavior of the Kerr spectra suggests a broadening of the optical response and a sensitivity to pairing in the intermediate region, especially in proximity to the plasma edge~\cite{taylor2013anomalous}.
Interestingly, in region III, the peak of the Kerr angle shifts towards higher energy for the nonchiral pairings.
In fact, for the nonchiral pairings, the Kerr angle is primarily dominated by the real part  of the frequency-dependent Hall conductivity, however for 
the chiral pairing $p+{\mi} p$, the Kerr angle receives contributions from both the real and imaginary parts, as indicated in Table.~\ref{table:1}.

%
\begin{table}[b]
\caption{Comparing the results obtained for maximum Kerr rotation (at region III) for three considered pairing symmetries: $p+{\mi} p$, $s^\prime+{\mi} d$, and $s^\prime+{\mi} d$.
 Polar Kerr spectrum is peaked at $\omega\approx 0.85$eV and $\omega \approx 0.95$eV for the chiral and nonchiral pairing, respectively; indicating by the vertical dashed lines in Fig.~\ref{fig3}.
Each data-set within individual cells corresponds to the calculated results for $\lambda=0$, $\lambda= g/2$, and $\lambda= g$, respectively.
}
\vspace{0.2cm}	
\centering
\begin{tabular}{|c|c|c|c|}
\hline
Pairings & $\theta^{\rm Kerr} 
$(nrad)
&
$\Im [\sigma^H] (10^{-8} e^2/\hbar)$
&
$\Re [\sigma^H] (10^{-8} e^2/\hbar)$
\\
 \hline
$p+{\mi}p$
& $ 25^{}, 26^{}, 54^{}$
& -6.2, -7, -10      & -7.3, -7.6, -18.4       \\
 \hline
$s^\prime+{\mi}d$
& $ 51^{}, 53^{}, 64^{}$
& 0.04, 0.042, 0.05      & -18.4, -19, -23.2         \\
 \hline
$s^\prime+{\mi}p$
& $ 42^{}, 44^{}, 53^{}$      & 0.033, 0.035, 0.043      & -15.2, -15, -19.3        \\
 \hline
 \end{tabular}
 \label{table:1}
\end{table}
%

%
Notably, the inclusion of spin-orbit coupling leads to an enhancement in the Kerr angle. This effect could be attributed to the introduction of spin magnetic order, which contributes to the observed finite Kerr signal in a spin system. Furthermore, the spin-orbit interaction induces interorbital mixing along the Brillouin zone diagonal.
At the end, in comparison with existing  experimental findings~\cite{katsufuji1996plane}, where $\theta^{\rm Kerr}_{\rm Exp}\approx 65$nrad was measured at $\omega=0.8$eV, the results obtained for all three pairings appear to be in reasonable agreement, particularly for the chiral order parameter.

\section{Summary}
\label{Sec:Summary}
We have presented a detailed theoretical study of the intrinsic Kerr response for the three-orbital model in the presence of spin-orbit coupling, $\lambda$, as an application model for Sr$_2$RuO$_4$ superconductor.
Our investigation has unveiled that the spontaneous Hall transport, resulting in a nonzero Kerr angle and finite Berry curvature, is primarily instigated by the quasi-1D orbitals $d_{xz}$ and $d_{yz}$ with a interorbital coupling, $g$.
We have demonstrated that the terms associated with the quasi-1D orbitals primarily influence the results, as they are directly proportional to $ g$. In contrast, the contribution of the orbital $d_{xy}$ to the optical Hall response  are expressed as a proportionality to  $\lambda^2g$, and can be neglected.
As a result, the absence of edge currents can be rationalized by considering the $d_{xz}$ and $d_{yz}$ orbitals as the primary active components. This choice arises because the components of the pairings in these orbitals are nearly decoupled, except in small regions of $\bk$-space where the Fermi surfaces closely intersect. Consequently, this scenario leads to a substantial reduction in edge currents.
The SOC plays a pivotal role in amplifying the intensity of the resonance peak, particularly when the frequency of electromagnetic radiation aligns with the characteristic scale of hybridization. This heightened hybridization, in turn, leads to a remarkable enhancement of the Kerr angle for possible  pairing states characterized by the breaking of time reversal symmetry. This enhancement can be attributed to the incorporation of contributions from spin magnetic order into the optical Hall response and the influence of Berry curvature.
%

	
	\section*{Acknowledgments}
	We acknowledge Thomas Scaffidi and Yunkyu Bang for the helpful discussions.

%
%
\appendix
\label{appendix}

\section{}
\label{coefficient}
Here, we offer a synophis of the numerator prefactors associated of each term in Eq.~(\ref{hall:conduc}), categorized by velocity type, while  neglecting   terms of order  ${\cal O}  \lambda^2g$.
For the terms include
$v_{x}^{yz}v_{y}^{xz-yz }$
and
$v_{y}^{yz}v_{x}^{xz-yz}$,
we can write
\be
\bl\no
&
4{\rm g}_{\bk}
(z-{z^\prime})
(z+{z^\prime})^2
\Im[
\Delta^{xz*}_{\bk}\Delta^{yz}_{\bk}
]
\\
&
\times
(z^2-E^{xz 2}_{\bk } )
(z^2-E^{yz 2}_{\bk } )
(z^2-E^{xy 2}_{\bk } )^2
\\
&
\times
({z^\prime}^2-E^{xz 2}_{\bk } )
({z^\prime}^2-E^{yz 2}_{\bk } )
({z^\prime}^2-E^{xy 2}_{\bk } )^2
+
{\cal O} \lambda^2 g
,
\el
\ee
and for the terms that include
$v_{x}^{xz}v_{y}^{xz-yz}$
and
$v_{y}^{xz}v_{x}^{xz-yz}, 
$
we have
\be
\bl\no
&
-4{\rm g}_{\bk}
(z-{z^\prime})
(z+{z^\prime})^2
\Im[
\Delta^{xz*}_{\bk}\Delta^{yz}_{\bk}
]
\\
&
\times
(z^2-E^{yz2}_{\bk })
(z^2-E^{xz2}_{\bk })
(z^2-E^{xy2}_{\bk })^2
\\
&
\times
({z^\prime}^2-E^{yz 2}_{\bk } )
({z^\prime}^2-E^{xz 2}_{\bk } )
({z^\prime}^2-E^{xy 2}_{\bk } )^2
+{\cal O} \lambda^2 g
.
\el
\ee
%
Furthermore, the inclusion of the following terms:
$
v_{x}^{yz}v_{y}^{xy}
$,
$v_{y}^{yz}v_{x}^{xy}$,
$v_{x}^{xz}v_{y}^{xy}$,
$v_{y}^{xz}v_{x}^{xy}$,
$v_{x}^{xy}v_{y}^{xz-yz}$, and
$v_{y}^{xy}v_{x}^{xz-yz}$,
 all are proportional to ${\cal O}\lambda^2g$, making them negligible.
For the rest of the components, these coefficients vanish.
%
%
%

%
%
\twocolumngrid
\bibliography{References}

\begin{thebibliography}{59}%
\makeatletter
\providecommand \@ifxundefined [1]{%
 \@ifx{#1\undefined}
}%
\providecommand \@ifnum [1]{%
 \ifnum #1\expandafter \@firstoftwo
 \else \expandafter \@secondoftwo
 \fi
}%
\providecommand \@ifx [1]{%
 \ifx #1\expandafter \@firstoftwo
 \else \expandafter \@secondoftwo
 \fi
}%
\providecommand \natexlab [1]{#1}%
\providecommand \enquote  [1]{``#1''}%
\providecommand \bibnamefont  [1]{#1}%
\providecommand \bibfnamefont [1]{#1}%
\providecommand \citenamefont [1]{#1}%
\providecommand \href@noop [0]{\@secondoftwo}%
\providecommand \href [0]{\begingroup \@sanitize@url \@href}%
\providecommand \@href[1]{\@@startlink{#1}\@@href}%
\providecommand \@@href[1]{\endgroup#1\@@endlink}%
\providecommand \@sanitize@url [0]{\catcode `\\12\catcode `\$12\catcode
  `\&12\catcode `\#12\catcode `\^12\catcode `\_12\catcode `\%12\relax}%
\providecommand \@@startlink[1]{}%
\providecommand \@@endlink[0]{}%
\providecommand \url  [0]{\begingroup\@sanitize@url \@url }%
\providecommand \@url [1]{\endgroup\@href {#1}{\urlprefix }}%
\providecommand \urlprefix  [0]{URL }%
\providecommand \Eprint [0]{\href }%
\providecommand \doibase [0]{https://doi.org/}%
\providecommand \selectlanguage [0]{\@gobble}%
\providecommand \bibinfo  [0]{\@secondoftwo}%
\providecommand \bibfield  [0]{\@secondoftwo}%
\providecommand \translation [1]{[#1]}%
\providecommand \BibitemOpen [0]{}%
\providecommand \bibitemStop [0]{}%
\providecommand \bibitemNoStop [0]{.\EOS\space}%
\providecommand \EOS [0]{\spacefactor3000\relax}%
\providecommand \BibitemShut  [1]{\csname bibitem#1\endcsname}%
\let\auto@bib@innerbib\@empty
\bibitem [{\citenamefont {Benhabib}\ \emph {et~al.}(2021)\citenamefont
  {Benhabib}, \citenamefont {Lupien}, \citenamefont {Paul}, \citenamefont
  {Berges}, \citenamefont {Dion}, \citenamefont {Nardone}, \citenamefont
  {Zitouni}, \citenamefont {Mao}, \citenamefont {Maeno}, \citenamefont
  {Georges}, \citenamefont {Taillefer},\ and\ \citenamefont
  {Proust}}]{benhabib2021ultrasound}%
  \BibitemOpen
  \bibfield  {author} {\bibinfo {author} {\bibfnamefont {S.}~\bibnamefont
  {Benhabib}}, \bibinfo {author} {\bibfnamefont {C.}~\bibnamefont {Lupien}},
  \bibinfo {author} {\bibfnamefont {I.}~\bibnamefont {Paul}}, \bibinfo {author}
  {\bibfnamefont {L.}~\bibnamefont {Berges}}, \bibinfo {author} {\bibfnamefont
  {M.}~\bibnamefont {Dion}}, \bibinfo {author} {\bibfnamefont {M.}~\bibnamefont
  {Nardone}}, \bibinfo {author} {\bibfnamefont {A.}~\bibnamefont {Zitouni}},
  \bibinfo {author} {\bibfnamefont {Z.~Q.}\ \bibnamefont {Mao}}, \bibinfo
  {author} {\bibfnamefont {Y.}~\bibnamefont {Maeno}}, \bibinfo {author}
  {\bibfnamefont {A.}~\bibnamefont {Georges}}, \bibinfo {author} {\bibfnamefont
  {L.}~\bibnamefont {Taillefer}},\ and\ \bibinfo {author} {\bibfnamefont
  {C.}~\bibnamefont {Proust}},\ }\bibfield  {title} {\bibinfo {title}
  {Ultrasound evidence for a two-component superconducting order parameter in
  {{${\mathrm{Sr}}_{2}{\mathrm{RuO}}_{4}$} }},\ }\href
  {https://doi.org/10.1038/s41567-020-1033-3} {\bibfield  {journal} {\bibinfo
  {journal} {Nature Physics}\ }\textbf {\bibinfo {volume} {17}},\ \bibinfo
  {pages} {194} (\bibinfo {year} {2021})}\BibitemShut {NoStop}%
\bibitem [{\citenamefont {Chronister}\ \emph {et~al.}(2021)\citenamefont
  {Chronister}, \citenamefont {Pustogow}, \citenamefont {Kikugawa},
  \citenamefont {Sokolov}, \citenamefont {Jerzembeck}, \citenamefont {Hicks},
  \citenamefont {Mackenzie}, \citenamefont {Bauer},\ and\ \citenamefont
  {Brown}}]{chronister2021evidence}%
  \BibitemOpen
  \bibfield  {author} {\bibinfo {author} {\bibfnamefont {A.}~\bibnamefont
  {Chronister}}, \bibinfo {author} {\bibfnamefont {A.}~\bibnamefont
  {Pustogow}}, \bibinfo {author} {\bibfnamefont {N.}~\bibnamefont {Kikugawa}},
  \bibinfo {author} {\bibfnamefont {D.~A.}\ \bibnamefont {Sokolov}}, \bibinfo
  {author} {\bibfnamefont {F.}~\bibnamefont {Jerzembeck}}, \bibinfo {author}
  {\bibfnamefont {C.~W.}\ \bibnamefont {Hicks}}, \bibinfo {author}
  {\bibfnamefont {A.~P.}\ \bibnamefont {Mackenzie}}, \bibinfo {author}
  {\bibfnamefont {E.~D.}\ \bibnamefont {Bauer}},\ and\ \bibinfo {author}
  {\bibfnamefont {S.~E.}\ \bibnamefont {Brown}},\ }\bibfield  {title} {\bibinfo
  {title} {Evidence for even parity unconventional superconductivity in
  {{${\mathrm{Sr}}_{2}{\mathrm{RuO}}_{4}$} }},\ }\href
  {https://doi.org/10.1073/pnas.2025313118} {\bibfield  {journal} {\bibinfo
  {journal} {Proceedings of the National Academy of Sciences}\ }\textbf
  {\bibinfo {volume} {118}},\ \bibinfo {pages} {e2025313118} (\bibinfo {year}
  {2021})}\BibitemShut {NoStop}%
\bibitem [{\citenamefont {Ghosh}\ \emph {et~al.}(2021)\citenamefont {Ghosh},
  \citenamefont {Shekhter}, \citenamefont {Jerzembeck}, \citenamefont
  {Kikugawa}, \citenamefont {Sokolov}, \citenamefont {Brando}, \citenamefont
  {Mackenzie}, \citenamefont {Hicks},\ and\ \citenamefont
  {Ramshaw}}]{ghosh2021thermodynamic}%
  \BibitemOpen
  \bibfield  {author} {\bibinfo {author} {\bibfnamefont {S.}~\bibnamefont
  {Ghosh}}, \bibinfo {author} {\bibfnamefont {A.}~\bibnamefont {Shekhter}},
  \bibinfo {author} {\bibfnamefont {F.}~\bibnamefont {Jerzembeck}}, \bibinfo
  {author} {\bibfnamefont {N.}~\bibnamefont {Kikugawa}}, \bibinfo {author}
  {\bibfnamefont {D.~A.}\ \bibnamefont {Sokolov}}, \bibinfo {author}
  {\bibfnamefont {M.}~\bibnamefont {Brando}}, \bibinfo {author} {\bibfnamefont
  {A.~P.}\ \bibnamefont {Mackenzie}}, \bibinfo {author} {\bibfnamefont {C.~W.}\
  \bibnamefont {Hicks}},\ and\ \bibinfo {author} {\bibfnamefont {B.~J.}\
  \bibnamefont {Ramshaw}},\ }\bibfield  {title} {\bibinfo {title}
  {Thermodynamic evidence for a two-component superconducting order parameter
  in {Sr$_2$RuO$_4$}},\ }\href {https://doi.org/10.1038/s41567-020-1032-4}
  {\bibfield  {journal} {\bibinfo  {journal} {Nature Physics}\ }\textbf
  {\bibinfo {volume} {17}},\ \bibinfo {pages} {199} (\bibinfo {year}
  {2021})}\BibitemShut {NoStop}%
\bibitem [{\citenamefont {Grinenko}\ \emph
  {et~al.}(2021{\natexlab{a}})\citenamefont {Grinenko}, \citenamefont {Ghosh},
  \citenamefont {Sarkar}, \citenamefont {Orain}, \citenamefont {Nikitin},
  \citenamefont {Elender}, \citenamefont {Das}, \citenamefont {Guguchia},
  \citenamefont {Br{\"u}ckner}, \citenamefont {Barber}, \citenamefont {Park},
  \citenamefont {Kikugawa}, \citenamefont {Sokolov}, \citenamefont {Bobowski},
  \citenamefont {Miyoshi}, \citenamefont {Maeno}, \citenamefont {Mackenzie},
  \citenamefont {Luetkens}, \citenamefont {Hicks},\ and\ \citenamefont
  {Klauss}}]{grinenko2021split}%
  \BibitemOpen
  \bibfield  {author} {\bibinfo {author} {\bibfnamefont {V.}~\bibnamefont
  {Grinenko}}, \bibinfo {author} {\bibfnamefont {S.}~\bibnamefont {Ghosh}},
  \bibinfo {author} {\bibfnamefont {R.}~\bibnamefont {Sarkar}}, \bibinfo
  {author} {\bibfnamefont {J.-C.}\ \bibnamefont {Orain}}, \bibinfo {author}
  {\bibfnamefont {A.}~\bibnamefont {Nikitin}}, \bibinfo {author} {\bibfnamefont
  {M.}~\bibnamefont {Elender}}, \bibinfo {author} {\bibfnamefont
  {D.}~\bibnamefont {Das}}, \bibinfo {author} {\bibfnamefont {Z.}~\bibnamefont
  {Guguchia}}, \bibinfo {author} {\bibfnamefont {F.}~\bibnamefont
  {Br{\"u}ckner}}, \bibinfo {author} {\bibfnamefont {M.~E.}\ \bibnamefont
  {Barber}}, \bibinfo {author} {\bibfnamefont {J.}~\bibnamefont {Park}},
  \bibinfo {author} {\bibfnamefont {N.}~\bibnamefont {Kikugawa}}, \bibinfo
  {author} {\bibfnamefont {D.~A.}\ \bibnamefont {Sokolov}}, \bibinfo {author}
  {\bibfnamefont {J.~S.}\ \bibnamefont {Bobowski}}, \bibinfo {author}
  {\bibfnamefont {T.}~\bibnamefont {Miyoshi}}, \bibinfo {author} {\bibfnamefont
  {Y.}~\bibnamefont {Maeno}}, \bibinfo {author} {\bibfnamefont {A.~P.}\
  \bibnamefont {Mackenzie}}, \bibinfo {author} {\bibfnamefont {H.}~\bibnamefont
  {Luetkens}}, \bibinfo {author} {\bibfnamefont {C.~W.}\ \bibnamefont
  {Hicks}},\ and\ \bibinfo {author} {\bibfnamefont {H.-H.}\ \bibnamefont
  {Klauss}},\ }\bibfield  {title} {\bibinfo {title} {Split superconducting and
  time-reversal symmetry-breaking transitions in
  {{${\mathrm{Sr}}_{2}{\mathrm{RuO}}_{4}$} } under stress},\ }\href
  {https://doi.org/10.1038/s41567-021-01182-7} {\bibfield  {journal} {\bibinfo
  {journal} {Nature Physics}\ }\textbf {\bibinfo {volume} {17}},\ \bibinfo
  {pages} {748} (\bibinfo {year} {2021}{\natexlab{a}})}\BibitemShut {NoStop}%
\bibitem [{\citenamefont {Grinenko}\ \emph
  {et~al.}(2021{\natexlab{b}})\citenamefont {Grinenko}, \citenamefont {Weston},
  \citenamefont {Caglieris}, \citenamefont {Wuttke}, \citenamefont {Hess},
  \citenamefont {Gottschall}, \citenamefont {Maccari}, \citenamefont
  {Gorbunov}, \citenamefont {Zherlitsyn}, \citenamefont {Wosnitza},
  \citenamefont {Rydh}, \citenamefont {Kihou}, \citenamefont {Lee},
  \citenamefont {Sarkar}, \citenamefont {Dengre}, \citenamefont {Garaud},
  \citenamefont {Charnukha}, \citenamefont {H{\"u}hne}, \citenamefont
  {Nielsch}, \citenamefont {B{\"u}chner}, \citenamefont {Klauss},\ and\
  \citenamefont {Babaev}}]{grinenko2021state}%
  \BibitemOpen
  \bibfield  {author} {\bibinfo {author} {\bibfnamefont {V.}~\bibnamefont
  {Grinenko}}, \bibinfo {author} {\bibfnamefont {D.}~\bibnamefont {Weston}},
  \bibinfo {author} {\bibfnamefont {F.}~\bibnamefont {Caglieris}}, \bibinfo
  {author} {\bibfnamefont {C.}~\bibnamefont {Wuttke}}, \bibinfo {author}
  {\bibfnamefont {C.}~\bibnamefont {Hess}}, \bibinfo {author} {\bibfnamefont
  {T.}~\bibnamefont {Gottschall}}, \bibinfo {author} {\bibfnamefont
  {I.}~\bibnamefont {Maccari}}, \bibinfo {author} {\bibfnamefont
  {D.}~\bibnamefont {Gorbunov}}, \bibinfo {author} {\bibfnamefont
  {S.}~\bibnamefont {Zherlitsyn}}, \bibinfo {author} {\bibfnamefont
  {J.}~\bibnamefont {Wosnitza}}, \bibinfo {author} {\bibfnamefont
  {A.}~\bibnamefont {Rydh}}, \bibinfo {author} {\bibfnamefont {K.}~\bibnamefont
  {Kihou}}, \bibinfo {author} {\bibfnamefont {C.-H.}\ \bibnamefont {Lee}},
  \bibinfo {author} {\bibfnamefont {R.}~\bibnamefont {Sarkar}}, \bibinfo
  {author} {\bibfnamefont {S.}~\bibnamefont {Dengre}}, \bibinfo {author}
  {\bibfnamefont {J.}~\bibnamefont {Garaud}}, \bibinfo {author} {\bibfnamefont
  {A.}~\bibnamefont {Charnukha}}, \bibinfo {author} {\bibfnamefont
  {R.}~\bibnamefont {H{\"u}hne}}, \bibinfo {author} {\bibfnamefont
  {K.}~\bibnamefont {Nielsch}}, \bibinfo {author} {\bibfnamefont
  {B.}~\bibnamefont {B{\"u}chner}}, \bibinfo {author} {\bibfnamefont {H.-H.}\
  \bibnamefont {Klauss}},\ and\ \bibinfo {author} {\bibfnamefont
  {E.}~\bibnamefont {Babaev}},\ }\bibfield  {title} {\bibinfo {title} {State
  with spontaneously broken time-reversal symmetry above the superconducting
  phase transition},\ }\href {https://doi.org/10.1038/s41567-021-01350-9}
  {\bibfield  {journal} {\bibinfo  {journal} {Nature Physics}\ }\textbf
  {\bibinfo {volume} {17}},\ \bibinfo {pages} {1254} (\bibinfo {year}
  {2021}{\natexlab{b}})}\BibitemShut {NoStop}%
\bibitem [{\citenamefont {Kivelson}\ \emph {et~al.}(2020)\citenamefont
  {Kivelson}, \citenamefont {Yuan}, \citenamefont {Ramshaw},\ and\
  \citenamefont {Thomale}}]{kivelson2020proposal}%
  \BibitemOpen
  \bibfield  {author} {\bibinfo {author} {\bibfnamefont {S.~A.}\ \bibnamefont
  {Kivelson}}, \bibinfo {author} {\bibfnamefont {A.~C.}\ \bibnamefont {Yuan}},
  \bibinfo {author} {\bibfnamefont {B.}~\bibnamefont {Ramshaw}},\ and\ \bibinfo
  {author} {\bibfnamefont {R.}~\bibnamefont {Thomale}},\ }\bibfield  {title}
  {\bibinfo {title} {A proposal for reconciling diverse experiments on the
  superconducting state in {{${\mathrm{Sr}}_{2}{\mathrm{RuO}}_{4}$} }},\ }\href
  {https://doi.org/10.1038/s41535-020-0245-1} {\bibfield  {journal} {\bibinfo
  {journal} {npj Quantum Materials}\ }\textbf {\bibinfo {volume} {5}},\
  \bibinfo {pages} {43} (\bibinfo {year} {2020})}\BibitemShut {NoStop}%
\bibitem [{\citenamefont {R\o{}mer}\ \emph {et~al.}(2021)\citenamefont
  {R\o{}mer}, \citenamefont {Hirschfeld},\ and\ \citenamefont
  {Andersen}}]{romer2021superconducting}%
  \BibitemOpen
  \bibfield  {author} {\bibinfo {author} {\bibfnamefont {A.~T.}\ \bibnamefont
  {R\o{}mer}}, \bibinfo {author} {\bibfnamefont {P.~J.}\ \bibnamefont
  {Hirschfeld}},\ and\ \bibinfo {author} {\bibfnamefont {B.~M.}\ \bibnamefont
  {Andersen}},\ }\bibfield  {title} {\bibinfo {title} {Superconducting state of
  {${\mathrm{Sr}}_{2}{\mathrm{RuO}}_{4}$ } in the presence of longer-range
  coulomb interactions},\ }\href {https://doi.org/10.1103/PhysRevB.104.064507}
  {\bibfield  {journal} {\bibinfo  {journal} {Phys. Rev. B}\ }\textbf {\bibinfo
  {volume} {104}},\ \bibinfo {pages} {064507} (\bibinfo {year}
  {2021})}\BibitemShut {NoStop}%
\bibitem [{\citenamefont {Bhattacharyya}\ \emph {et~al.}(2023)\citenamefont
  {Bhattacharyya}, \citenamefont {Kreisel}, \citenamefont {Kong}, \citenamefont
  {Berlijn}, \citenamefont {R\o{}mer}, \citenamefont {Andersen},\ and\
  \citenamefont {Hirschfeld}}]{Bhattacharyya2023Superconducting}%
  \BibitemOpen
  \bibfield  {author} {\bibinfo {author} {\bibfnamefont {S.}~\bibnamefont
  {Bhattacharyya}}, \bibinfo {author} {\bibfnamefont {A.}~\bibnamefont
  {Kreisel}}, \bibinfo {author} {\bibfnamefont {X.}~\bibnamefont {Kong}},
  \bibinfo {author} {\bibfnamefont {T.}~\bibnamefont {Berlijn}}, \bibinfo
  {author} {\bibfnamefont {A.~T.}\ \bibnamefont {R\o{}mer}}, \bibinfo {author}
  {\bibfnamefont {B.~M.}\ \bibnamefont {Andersen}},\ and\ \bibinfo {author}
  {\bibfnamefont {P.~J.}\ \bibnamefont {Hirschfeld}},\ }\bibfield  {title}
  {\bibinfo {title} {Superconducting gap symmetry from bogoliubov quasiparticle
  interference analysis on {${\mathrm{Sr}}_{2}{\mathrm{RuO}}_{4}$}},\ }\href
  {https://doi.org/10.1103/PhysRevB.107.144505} {\bibfield  {journal} {\bibinfo
   {journal} {Phys. Rev. B}\ }\textbf {\bibinfo {volume} {107}},\ \bibinfo
  {pages} {144505} (\bibinfo {year} {2023})}\BibitemShut {NoStop}%
\bibitem [{\citenamefont {K\"aser}\ \emph {et~al.}(2022)\citenamefont
  {K\"aser}, \citenamefont {Strand}, \citenamefont {Wentzell}, \citenamefont
  {Georges}, \citenamefont {Parcollet},\ and\ \citenamefont
  {Hansmann}}]{Kaser2022Interorbital}%
  \BibitemOpen
  \bibfield  {author} {\bibinfo {author} {\bibfnamefont {S.}~\bibnamefont
  {K\"aser}}, \bibinfo {author} {\bibfnamefont {H.~U.~R.}\ \bibnamefont
  {Strand}}, \bibinfo {author} {\bibfnamefont {N.}~\bibnamefont {Wentzell}},
  \bibinfo {author} {\bibfnamefont {A.}~\bibnamefont {Georges}}, \bibinfo
  {author} {\bibfnamefont {O.}~\bibnamefont {Parcollet}},\ and\ \bibinfo
  {author} {\bibfnamefont {P.}~\bibnamefont {Hansmann}},\ }\bibfield  {title}
  {\bibinfo {title} {Interorbital singlet pairing in
  {${\mathrm{Sr}}_{2}{\mathrm{RuO}}_{4}$}: A hund's superconductor},\ }\href
  {https://doi.org/10.1103/PhysRevB.105.155101} {\bibfield  {journal} {\bibinfo
   {journal} {Phys. Rev. B}\ }\textbf {\bibinfo {volume} {105}},\ \bibinfo
  {pages} {155101} (\bibinfo {year} {2022})}\BibitemShut {NoStop}%
\bibitem [{\citenamefont {Ikegaya}\ \emph {et~al.}(2020)\citenamefont
  {Ikegaya}, \citenamefont {Yada}, \citenamefont {Tanaka}, \citenamefont
  {Kashiwaya}, \citenamefont {Asano},\ and\ \citenamefont
  {Manske}}]{Ikegaya2020Identification}%
  \BibitemOpen
  \bibfield  {author} {\bibinfo {author} {\bibfnamefont {S.}~\bibnamefont
  {Ikegaya}}, \bibinfo {author} {\bibfnamefont {K.}~\bibnamefont {Yada}},
  \bibinfo {author} {\bibfnamefont {Y.}~\bibnamefont {Tanaka}}, \bibinfo
  {author} {\bibfnamefont {S.}~\bibnamefont {Kashiwaya}}, \bibinfo {author}
  {\bibfnamefont {Y.}~\bibnamefont {Asano}},\ and\ \bibinfo {author}
  {\bibfnamefont {D.}~\bibnamefont {Manske}},\ }\bibfield  {title} {\bibinfo
  {title} {Identification of spin-triplet superconductivity through a
  helical-chiral phase transition in {${\mathrm{Sr}}_{2}{\mathrm{RuO}}_{4}$}
  thin films},\ }\href {https://doi.org/10.1103/PhysRevB.101.220501} {\bibfield
   {journal} {\bibinfo  {journal} {Phys. Rev. B}\ }\textbf {\bibinfo {volume}
  {101}},\ \bibinfo {pages} {220501} (\bibinfo {year} {2020})}\BibitemShut
  {NoStop}%
\bibitem [{\citenamefont {Gingras}\ \emph {et~al.}(2022)\citenamefont
  {Gingras}, \citenamefont {Allaglo}, \citenamefont {Nourafkan}, \citenamefont
  {C\^ot\'e},\ and\ \citenamefont {Tremblay}}]{Gingras2022Superconductivity}%
  \BibitemOpen
  \bibfield  {author} {\bibinfo {author} {\bibfnamefont {O.}~\bibnamefont
  {Gingras}}, \bibinfo {author} {\bibfnamefont {N.}~\bibnamefont {Allaglo}},
  \bibinfo {author} {\bibfnamefont {R.}~\bibnamefont {Nourafkan}}, \bibinfo
  {author} {\bibfnamefont {M.}~\bibnamefont {C\^ot\'e}},\ and\ \bibinfo
  {author} {\bibfnamefont {A.-M.~S.}\ \bibnamefont {Tremblay}},\ }\bibfield
  {title} {\bibinfo {title} {Superconductivity in correlated multiorbital
  systems with spin-orbit coupling: Coexistence of even- and odd-frequency
  pairing, and the case of {${\mathrm{Sr}}_{2}{\mathrm{RuO}}_{4}$}},\ }\href
  {https://doi.org/10.1103/PhysRevB.106.064513} {\bibfield  {journal} {\bibinfo
   {journal} {Phys. Rev. B}\ }\textbf {\bibinfo {volume} {106}},\ \bibinfo
  {pages} {064513} (\bibinfo {year} {2022})}\BibitemShut {NoStop}%
\bibitem [{\citenamefont {Wagner}\ \emph {et~al.}(2021)\citenamefont {Wagner},
  \citenamefont {R\o{}ising}, \citenamefont {Flicker},\ and\ \citenamefont
  {Simon}}]{Wagner2021Microscopic}%
  \BibitemOpen
  \bibfield  {author} {\bibinfo {author} {\bibfnamefont {G.}~\bibnamefont
  {Wagner}}, \bibinfo {author} {\bibfnamefont {H.~S.}\ \bibnamefont
  {R\o{}ising}}, \bibinfo {author} {\bibfnamefont {F.}~\bibnamefont
  {Flicker}},\ and\ \bibinfo {author} {\bibfnamefont {S.~H.}\ \bibnamefont
  {Simon}},\ }\bibfield  {title} {\bibinfo {title} {Microscopic ginzburg-landau
  theory and singlet ordering in {${\mathrm{Sr}}_{2}{\mathrm{RuO}}_{4}$}},\
  }\href {https://doi.org/10.1103/PhysRevB.104.134506} {\bibfield  {journal}
  {\bibinfo  {journal} {Phys. Rev. B}\ }\textbf {\bibinfo {volume} {104}},\
  \bibinfo {pages} {134506} (\bibinfo {year} {2021})}\BibitemShut {NoStop}%
\bibitem [{\citenamefont {Lindquist}\ and\ \citenamefont
  {Kee}(2020)}]{Lindquist2020Distinct}%
  \BibitemOpen
  \bibfield  {author} {\bibinfo {author} {\bibfnamefont {A.~W.}\ \bibnamefont
  {Lindquist}}\ and\ \bibinfo {author} {\bibfnamefont {H.-Y.}\ \bibnamefont
  {Kee}},\ }\bibfield  {title} {\bibinfo {title} {Distinct reduction of knight
  shift in superconducting state of {${\mathrm{Sr}}_{2}{\mathrm{RuO}}_{4}$}
  under uniaxial strain},\ }\href
  {https://doi.org/10.1103/PhysRevResearch.2.032055} {\bibfield  {journal}
  {\bibinfo  {journal} {Phys. Rev. Res.}\ }\textbf {\bibinfo {volume} {2}},\
  \bibinfo {pages} {032055} (\bibinfo {year} {2020})}\BibitemShut {NoStop}%
\bibitem [{\citenamefont {Lindquist}\ and\ \citenamefont
  {Kee}(2023)}]{Lindquist2023Reconciling}%
  \BibitemOpen
  \bibfield  {author} {\bibinfo {author} {\bibfnamefont {A.~W.}\ \bibnamefont
  {Lindquist}}\ and\ \bibinfo {author} {\bibfnamefont {H.-Y.}\ \bibnamefont
  {Kee}},\ }\bibfield  {title} {\bibinfo {title} {Reconciling the
  {$\ensuremath{\pi}$} phase shift in josephson junction experiments with
  even-parity superconductivity in {${\mathrm{Sr}}_{2}{\mathrm{RuO}}_{4}$}},\
  }\href {https://doi.org/10.1103/PhysRevB.107.014506} {\bibfield  {journal}
  {\bibinfo  {journal} {Phys. Rev. B}\ }\textbf {\bibinfo {volume} {107}},\
  \bibinfo {pages} {014506} (\bibinfo {year} {2023})}\BibitemShut {NoStop}%
\bibitem [{\citenamefont {Willa}\ \emph {et~al.}(2021)\citenamefont {Willa},
  \citenamefont {Hecker}, \citenamefont {Fernandes},\ and\ \citenamefont
  {Schmalian}}]{Willa2021Inhomogeneous}%
  \BibitemOpen
  \bibfield  {author} {\bibinfo {author} {\bibfnamefont {R.}~\bibnamefont
  {Willa}}, \bibinfo {author} {\bibfnamefont {M.}~\bibnamefont {Hecker}},
  \bibinfo {author} {\bibfnamefont {R.~M.}\ \bibnamefont {Fernandes}},\ and\
  \bibinfo {author} {\bibfnamefont {J.}~\bibnamefont {Schmalian}},\ }\bibfield
  {title} {\bibinfo {title} {Inhomogeneous time-reversal symmetry breaking in
  {${\mathrm{Sr}}_{2}{\mathrm{RuO}}_{4}$}},\ }\href
  {https://doi.org/10.1103/PhysRevB.104.024511} {\bibfield  {journal} {\bibinfo
   {journal} {Phys. Rev. B}\ }\textbf {\bibinfo {volume} {104}},\ \bibinfo
  {pages} {024511} (\bibinfo {year} {2021})}\BibitemShut {NoStop}%
\bibitem [{\citenamefont {R\o{}mer}\ \emph {et~al.}(2022)\citenamefont
  {R\o{}mer}, \citenamefont {Maier}, \citenamefont {Kreisel}, \citenamefont
  {Hirschfeld},\ and\ \citenamefont {Andersen}}]{Romer2022Leading}%
  \BibitemOpen
  \bibfield  {author} {\bibinfo {author} {\bibfnamefont {A.~T.}\ \bibnamefont
  {R\o{}mer}}, \bibinfo {author} {\bibfnamefont {T.~A.}\ \bibnamefont {Maier}},
  \bibinfo {author} {\bibfnamefont {A.}~\bibnamefont {Kreisel}}, \bibinfo
  {author} {\bibfnamefont {P.~J.}\ \bibnamefont {Hirschfeld}},\ and\ \bibinfo
  {author} {\bibfnamefont {B.~M.}\ \bibnamefont {Andersen}},\ }\bibfield
  {title} {\bibinfo {title} {Leading superconducting instabilities in
  three-dimensional models for {${\mathrm{Sr}}_{2}{\mathrm{RuO}}_{4}$}},\
  }\href {https://doi.org/10.1103/PhysRevResearch.4.033011} {\bibfield
  {journal} {\bibinfo  {journal} {Phys. Rev. Res.}\ }\textbf {\bibinfo {volume}
  {4}},\ \bibinfo {pages} {033011} (\bibinfo {year} {2022})}\BibitemShut
  {NoStop}%
\bibitem [{\citenamefont {Rice}\ and\ \citenamefont
  {Sigrist}(1995)}]{rice1995sr2ruo4}%
  \BibitemOpen
  \bibfield  {author} {\bibinfo {author} {\bibfnamefont {T.~M.}\ \bibnamefont
  {Rice}}\ and\ \bibinfo {author} {\bibfnamefont {M.}~\bibnamefont {Sigrist}},\
  }\bibfield  {title} {\bibinfo {title}
  {{${\mathrm{Sr}}_{2}{\mathrm{RuO}}_{4}$} : an electronic analogue of
  {3He}?},\ }\href {https://doi.org/10.1088/0953-8984/7/47/002} {\bibfield
  {journal} {\bibinfo  {journal} {Journal of Physics: Condensed Matter}\
  }\textbf {\bibinfo {volume} {7}},\ \bibinfo {pages} {L643} (\bibinfo {year}
  {1995})}\BibitemShut {NoStop}%
\bibitem [{\citenamefont {Luke}\ \emph {et~al.}(1998)\citenamefont {Luke},
  \citenamefont {Fudamoto}, \citenamefont {Kojima}, \citenamefont {Larkin},
  \citenamefont {Merrin}, \citenamefont {Nachumi}, \citenamefont {Uemura},
  \citenamefont {Maeno}, \citenamefont {Mao}, \citenamefont {Mori},
  \citenamefont {Nakamura},\ and\ \citenamefont {Sigrist}}]{luke1998time}%
  \BibitemOpen
  \bibfield  {author} {\bibinfo {author} {\bibfnamefont {G.~M.}\ \bibnamefont
  {Luke}}, \bibinfo {author} {\bibfnamefont {Y.}~\bibnamefont {Fudamoto}},
  \bibinfo {author} {\bibfnamefont {K.~M.}\ \bibnamefont {Kojima}}, \bibinfo
  {author} {\bibfnamefont {M.~I.}\ \bibnamefont {Larkin}}, \bibinfo {author}
  {\bibfnamefont {J.}~\bibnamefont {Merrin}}, \bibinfo {author} {\bibfnamefont
  {B.}~\bibnamefont {Nachumi}}, \bibinfo {author} {\bibfnamefont {Y.~J.}\
  \bibnamefont {Uemura}}, \bibinfo {author} {\bibfnamefont {Y.}~\bibnamefont
  {Maeno}}, \bibinfo {author} {\bibfnamefont {Z.~Q.}\ \bibnamefont {Mao}},
  \bibinfo {author} {\bibfnamefont {Y.}~\bibnamefont {Mori}}, \bibinfo {author}
  {\bibfnamefont {H.}~\bibnamefont {Nakamura}},\ and\ \bibinfo {author}
  {\bibfnamefont {M.}~\bibnamefont {Sigrist}},\ }\bibfield  {title} {\bibinfo
  {title} {Time-reversal symmetry-breaking superconductivity in
  {${\mathrm{Sr}}_{2}{\mathrm{RuO}}_{4}$}},\ }\href
  {https://doi.org/10.1038/29038} {\bibfield  {journal} {\bibinfo  {journal}
  {Nature}\ }\textbf {\bibinfo {volume} {394}},\ \bibinfo {pages} {558}
  (\bibinfo {year} {1998})}\BibitemShut {NoStop}%
\bibitem [{\citenamefont {Xia}\ \emph {et~al.}(2006)\citenamefont {Xia},
  \citenamefont {Maeno}, \citenamefont {Beyersdorf}, \citenamefont {Fejer},\
  and\ \citenamefont {Kapitulnik}}]{xia2006high}%
  \BibitemOpen
  \bibfield  {author} {\bibinfo {author} {\bibfnamefont {J.}~\bibnamefont
  {Xia}}, \bibinfo {author} {\bibfnamefont {Y.}~\bibnamefont {Maeno}}, \bibinfo
  {author} {\bibfnamefont {P.~T.}\ \bibnamefont {Beyersdorf}}, \bibinfo
  {author} {\bibfnamefont {M.~M.}\ \bibnamefont {Fejer}},\ and\ \bibinfo
  {author} {\bibfnamefont {A.}~\bibnamefont {Kapitulnik}},\ }\bibfield  {title}
  {\bibinfo {title} {High resolution polar kerr effect measurements of
  {${\mathrm{Sr}}_{2}{\mathrm{RuO}}_{4}$}: Evidence for broken time-reversal
  symmetry in the superconducting state},\ }\href
  {https://doi.org/10.1103/PhysRevLett.97.167002} {\bibfield  {journal}
  {\bibinfo  {journal} {Phys. Rev. Lett.}\ }\textbf {\bibinfo {volume} {97}},\
  \bibinfo {pages} {167002} (\bibinfo {year} {2006})}\BibitemShut {NoStop}%
\bibitem [{\citenamefont {Scaffidi}\ and\ \citenamefont
  {Simon}(2015)}]{scaffidi2015large}%
  \BibitemOpen
  \bibfield  {author} {\bibinfo {author} {\bibfnamefont {T.}~\bibnamefont
  {Scaffidi}}\ and\ \bibinfo {author} {\bibfnamefont {S.~H.}\ \bibnamefont
  {Simon}},\ }\bibfield  {title} {\bibinfo {title} {Large chern number and edge
  currents in {${\mathrm{Sr}}_{2}{\mathrm{RuO}}_{4}$}},\ }\href
  {https://doi.org/10.1103/PhysRevLett.115.087003} {\bibfield  {journal}
  {\bibinfo  {journal} {Phys. Rev. Lett.}\ }\textbf {\bibinfo {volume} {115}},\
  \bibinfo {pages} {087003} (\bibinfo {year} {2015})}\BibitemShut {NoStop}%
\bibitem [{\citenamefont {Hicks}\ \emph {et~al.}(2014)\citenamefont {Hicks},
  \citenamefont {Brodsky}, \citenamefont {Yelland}, \citenamefont {Gibbs},
  \citenamefont {Bruin}, \citenamefont {Barber}, \citenamefont {Edkins},
  \citenamefont {Nishimura}, \citenamefont {Yonezawa}, \citenamefont {Maeno}
  \emph {et~al.}}]{hicks2014strong}%
  \BibitemOpen
  \bibfield  {author} {\bibinfo {author} {\bibfnamefont {C.~W.}\ \bibnamefont
  {Hicks}}, \bibinfo {author} {\bibfnamefont {D.~O.}\ \bibnamefont {Brodsky}},
  \bibinfo {author} {\bibfnamefont {E.~A.}\ \bibnamefont {Yelland}}, \bibinfo
  {author} {\bibfnamefont {A.~S.}\ \bibnamefont {Gibbs}}, \bibinfo {author}
  {\bibfnamefont {J.~A.}\ \bibnamefont {Bruin}}, \bibinfo {author}
  {\bibfnamefont {M.~E.}\ \bibnamefont {Barber}}, \bibinfo {author}
  {\bibfnamefont {S.~D.}\ \bibnamefont {Edkins}}, \bibinfo {author}
  {\bibfnamefont {K.}~\bibnamefont {Nishimura}}, \bibinfo {author}
  {\bibfnamefont {S.}~\bibnamefont {Yonezawa}}, \bibinfo {author}
  {\bibfnamefont {Y.}~\bibnamefont {Maeno}}, \emph {et~al.},\ }\bibfield
  {title} {\bibinfo {title} {Strong increase of tc of
  {{${\mathrm{Sr}}_{2}{\mathrm{RuO}}_{4}$} } under both tensile and compressive
  strain},\ }\href {https://doi.org/10.1126/science.1248292} {\bibfield
  {journal} {\bibinfo  {journal} {Science}\ }\textbf {\bibinfo {volume}
  {344}},\ \bibinfo {pages} {283} (\bibinfo {year} {2014})}\BibitemShut
  {NoStop}%
\bibitem [{\citenamefont {Steppke}\ \emph {et~al.}(2017)\citenamefont
  {Steppke}, \citenamefont {Zhao}, \citenamefont {Barber}, \citenamefont
  {Scaffidi}, \citenamefont {Jerzembeck}, \citenamefont {Rosner}, \citenamefont
  {Gibbs}, \citenamefont {Maeno}, \citenamefont {Simon}, \citenamefont
  {Mackenzie},\ and\ \citenamefont {Hicks}}]{steppke2017strong}%
  \BibitemOpen
  \bibfield  {author} {\bibinfo {author} {\bibfnamefont {A.}~\bibnamefont
  {Steppke}}, \bibinfo {author} {\bibfnamefont {L.}~\bibnamefont {Zhao}},
  \bibinfo {author} {\bibfnamefont {M.~E.}\ \bibnamefont {Barber}}, \bibinfo
  {author} {\bibfnamefont {T.}~\bibnamefont {Scaffidi}}, \bibinfo {author}
  {\bibfnamefont {F.}~\bibnamefont {Jerzembeck}}, \bibinfo {author}
  {\bibfnamefont {H.}~\bibnamefont {Rosner}}, \bibinfo {author} {\bibfnamefont
  {A.~S.}\ \bibnamefont {Gibbs}}, \bibinfo {author} {\bibfnamefont
  {Y.}~\bibnamefont {Maeno}}, \bibinfo {author} {\bibfnamefont {S.~H.}\
  \bibnamefont {Simon}}, \bibinfo {author} {\bibfnamefont {A.~P.}\ \bibnamefont
  {Mackenzie}},\ and\ \bibinfo {author} {\bibfnamefont {C.~W.}\ \bibnamefont
  {Hicks}},\ }\bibfield  {title} {\bibinfo {title} {Strong peak in tc of
  {${\mathrm{Sr}}_{2}{\mathrm{RuO}}_{4}$} under uniaxial pressure},\ }\href
  {https://doi.org/10.1126/science.aaf9398} {\bibfield  {journal} {\bibinfo
  {journal} {Science}\ }\textbf {\bibinfo {volume} {355}},\ \bibinfo {pages}
  {eaaf9398} (\bibinfo {year} {2017})}\BibitemShut {NoStop}%
\bibitem [{\citenamefont {Ishida}\ \emph {et~al.}(2020)\citenamefont {Ishida},
  \citenamefont {Manago}, \citenamefont {Kinjo},\ and\ \citenamefont
  {Maeno}}]{ishida2020reduction}%
  \BibitemOpen
  \bibfield  {author} {\bibinfo {author} {\bibfnamefont {K.}~\bibnamefont
  {Ishida}}, \bibinfo {author} {\bibfnamefont {M.}~\bibnamefont {Manago}},
  \bibinfo {author} {\bibfnamefont {K.}~\bibnamefont {Kinjo}},\ and\ \bibinfo
  {author} {\bibfnamefont {Y.}~\bibnamefont {Maeno}},\ }\bibfield  {title}
  {\bibinfo {title} {Reduction of the 17o knight shift in the superconducting
  state and the heat-up effect by nmr pulses on
  {{${\mathrm{Sr}}_{2}{\mathrm{RuO}}_{4}$} }},\ }\href
  {https://doi.org/10.7566/JPSJ.89.034712} {\bibfield  {journal} {\bibinfo
  {journal} {Journal of the Physical Society of Japan}\ }\textbf {\bibinfo
  {volume} {89}},\ \bibinfo {pages} {034712} (\bibinfo {year}
  {2020})}\BibitemShut {NoStop}%
\bibitem [{\citenamefont {Agterberg}\ \emph {et~al.}(1997)\citenamefont
  {Agterberg}, \citenamefont {Rice},\ and\ \citenamefont
  {Sigrist}}]{agterberg1997orbital}%
  \BibitemOpen
  \bibfield  {author} {\bibinfo {author} {\bibfnamefont {D.~F.}\ \bibnamefont
  {Agterberg}}, \bibinfo {author} {\bibfnamefont {T.~M.}\ \bibnamefont
  {Rice}},\ and\ \bibinfo {author} {\bibfnamefont {M.}~\bibnamefont
  {Sigrist}},\ }\bibfield  {title} {\bibinfo {title} {Orbital dependent
  superconductivity in {${\mathrm{Sr}}_{2}{\mathrm{RuO}}_{4}$}},\ }\href
  {https://doi.org/10.1103/PhysRevLett.78.3374} {\bibfield  {journal} {\bibinfo
   {journal} {Phys. Rev. Lett.}\ }\textbf {\bibinfo {volume} {78}},\ \bibinfo
  {pages} {3374} (\bibinfo {year} {1997})}\BibitemShut {NoStop}%
\bibitem [{\citenamefont {Cobo}\ \emph {et~al.}(2016)\citenamefont {Cobo},
  \citenamefont {Ahn}, \citenamefont {Eremin},\ and\ \citenamefont
  {Akbari}}]{Cobo2016}%
  \BibitemOpen
  \bibfield  {author} {\bibinfo {author} {\bibfnamefont {S.}~\bibnamefont
  {Cobo}}, \bibinfo {author} {\bibfnamefont {F.}~\bibnamefont {Ahn}}, \bibinfo
  {author} {\bibfnamefont {I.}~\bibnamefont {Eremin}},\ and\ \bibinfo {author}
  {\bibfnamefont {A.}~\bibnamefont {Akbari}},\ }\bibfield  {title} {\bibinfo
  {title} {Anisotropic spin fluctuations in
  {${\mathrm{Sr}}_{2}{\mathrm{RuO}}_{4}$}: Role of spin-orbit coupling and
  induced strain},\ }\href {https://doi.org/10.1103/PhysRevB.94.224507}
  {\bibfield  {journal} {\bibinfo  {journal} {Phys. Rev. B}\ }\textbf {\bibinfo
  {volume} {94}},\ \bibinfo {pages} {224507} (\bibinfo {year}
  {2016})}\BibitemShut {NoStop}%
\bibitem [{\citenamefont {Barber}\ \emph {et~al.}(2018)\citenamefont {Barber},
  \citenamefont {Gibbs}, \citenamefont {Maeno}, \citenamefont {Mackenzie},\
  and\ \citenamefont {Hicks}}]{barber2018resistivity}%
  \BibitemOpen
  \bibfield  {author} {\bibinfo {author} {\bibfnamefont {M.~E.}\ \bibnamefont
  {Barber}}, \bibinfo {author} {\bibfnamefont {A.~S.}\ \bibnamefont {Gibbs}},
  \bibinfo {author} {\bibfnamefont {Y.}~\bibnamefont {Maeno}}, \bibinfo
  {author} {\bibfnamefont {A.~P.}\ \bibnamefont {Mackenzie}},\ and\ \bibinfo
  {author} {\bibfnamefont {C.~W.}\ \bibnamefont {Hicks}},\ }\bibfield  {title}
  {\bibinfo {title} {Resistivity in the vicinity of a van hove singularity:
  {${\mathrm{Sr}}_{2}{\mathrm{RuO}}_{4}$} under uniaxial pressure},\ }\href
  {https://doi.org/10.1103/PhysRevLett.120.076602} {\bibfield  {journal}
  {\bibinfo  {journal} {Phys. Rev. Lett.}\ }\textbf {\bibinfo {volume} {120}},\
  \bibinfo {pages} {076602} (\bibinfo {year} {2018})}\BibitemShut {NoStop}%
\bibitem [{\citenamefont {Yu}\ \emph {et~al.}(2020)\citenamefont {Yu},
  \citenamefont {Brown}, \citenamefont {Raghu},\ and\ \citenamefont
  {Yang}}]{Yu2020Critical}%
  \BibitemOpen
  \bibfield  {author} {\bibinfo {author} {\bibfnamefont {Y.}~\bibnamefont
  {Yu}}, \bibinfo {author} {\bibfnamefont {S.}~\bibnamefont {Brown}}, \bibinfo
  {author} {\bibfnamefont {S.}~\bibnamefont {Raghu}},\ and\ \bibinfo {author}
  {\bibfnamefont {K.}~\bibnamefont {Yang}},\ }\bibfield  {title} {\bibinfo
  {title} {Critical temperature ${T}_{c}$ and pauli limited critical field of
  {${\mathrm{Sr}}_{2}{\mathrm{RuO}}_{4}$}: Uniaxial strain dependence},\ }\href
  {https://doi.org/10.1103/PhysRevB.102.014509} {\bibfield  {journal} {\bibinfo
   {journal} {Phys. Rev. B}\ }\textbf {\bibinfo {volume} {102}},\ \bibinfo
  {pages} {014509} (\bibinfo {year} {2020})}\BibitemShut {NoStop}%
\bibitem [{\citenamefont {Kim}\ \emph {et~al.}(2023)\citenamefont {Kim},
  \citenamefont {Khmelevskyi}, \citenamefont {Franchini},\ and\ \citenamefont
  {Mazin}}]{Kim2023Suppressed}%
  \BibitemOpen
  \bibfield  {author} {\bibinfo {author} {\bibfnamefont {B.}~\bibnamefont
  {Kim}}, \bibinfo {author} {\bibfnamefont {S.}~\bibnamefont {Khmelevskyi}},
  \bibinfo {author} {\bibfnamefont {C.}~\bibnamefont {Franchini}},\ and\
  \bibinfo {author} {\bibfnamefont {I.~I.}\ \bibnamefont {Mazin}},\ }\bibfield
  {title} {\bibinfo {title} {Suppressed fluctuations as the origin of the
  static magnetic order in strained {${\mathrm{Sr}}_{2}{\mathrm{RuO}}_{4}$}},\
  }\href {https://doi.org/10.1103/PhysRevLett.130.026702} {\bibfield  {journal}
  {\bibinfo  {journal} {Phys. Rev. Lett.}\ }\textbf {\bibinfo {volume} {130}},\
  \bibinfo {pages} {026702} (\bibinfo {year} {2023})}\BibitemShut {NoStop}%
\bibitem [{\citenamefont {Robbins}\ \emph {et~al.}(2017)\citenamefont
  {Robbins}, \citenamefont {Annett},\ and\ \citenamefont
  {Gradhand}}]{robbins2017effect}%
  \BibitemOpen
  \bibfield  {author} {\bibinfo {author} {\bibfnamefont {J.}~\bibnamefont
  {Robbins}}, \bibinfo {author} {\bibfnamefont {J.~F.}\ \bibnamefont
  {Annett}},\ and\ \bibinfo {author} {\bibfnamefont {M.}~\bibnamefont
  {Gradhand}},\ }\bibfield  {title} {\bibinfo {title} {Effect of spin-orbit
  coupling on the polar kerr effect in
  {${\mathrm{Sr}}_{2}{\mathrm{RuO}}_{4}$}},\ }\href
  {https://doi.org/10.1103/PhysRevB.96.144503} {\bibfield  {journal} {\bibinfo
  {journal} {Phys. Rev. B}\ }\textbf {\bibinfo {volume} {96}},\ \bibinfo
  {pages} {144503} (\bibinfo {year} {2017})}\BibitemShut {NoStop}%
\bibitem [{\citenamefont {Gradhand}\ and\ \citenamefont
  {Annett}(2014)}]{Gradhand_2014}%
  \BibitemOpen
  \bibfield  {author} {\bibinfo {author} {\bibfnamefont {M.}~\bibnamefont
  {Gradhand}}\ and\ \bibinfo {author} {\bibfnamefont {J.~F.}\ \bibnamefont
  {Annett}},\ }\bibfield  {title} {\bibinfo {title} {The berry curvature of the
  bogoliubov quasiparticle bloch states in the unconventional superconductor
  {${\mathrm{Sr}}_{2}{\mathrm{RuO}}_{4}$}},\ }\href
  {https://doi.org/10.1088/0953-8984/26/27/274205} {\bibfield  {journal}
  {\bibinfo  {journal} {Journal of Physics: Condensed Matter}\ }\textbf
  {\bibinfo {volume} {26}},\ \bibinfo {pages} {274205} (\bibinfo {year}
  {2014})}\BibitemShut {NoStop}%
\bibitem [{\citenamefont {Sidis}\ \emph {et~al.}(1999)\citenamefont {Sidis},
  \citenamefont {Braden}, \citenamefont {Bourges}, \citenamefont {Hennion},
  \citenamefont {NishiZaki}, \citenamefont {Maeno},\ and\ \citenamefont
  {Mori}}]{Sidis1999Evidence}%
  \BibitemOpen
  \bibfield  {author} {\bibinfo {author} {\bibfnamefont {Y.}~\bibnamefont
  {Sidis}}, \bibinfo {author} {\bibfnamefont {M.}~\bibnamefont {Braden}},
  \bibinfo {author} {\bibfnamefont {P.}~\bibnamefont {Bourges}}, \bibinfo
  {author} {\bibfnamefont {B.}~\bibnamefont {Hennion}}, \bibinfo {author}
  {\bibfnamefont {S.}~\bibnamefont {NishiZaki}}, \bibinfo {author}
  {\bibfnamefont {Y.}~\bibnamefont {Maeno}},\ and\ \bibinfo {author}
  {\bibfnamefont {Y.}~\bibnamefont {Mori}},\ }\bibfield  {title} {\bibinfo
  {title} {Evidence for incommensurate spin fluctuations in
  {${\mathrm{Sr}}_{2}{\mathrm{RuO}}_{4}$}},\ }\href
  {https://doi.org/10.1103/PhysRevLett.83.3320} {\bibfield  {journal} {\bibinfo
   {journal} {Phys. Rev. Lett.}\ }\textbf {\bibinfo {volume} {83}},\ \bibinfo
  {pages} {3320} (\bibinfo {year} {1999})}\BibitemShut {NoStop}%
\bibitem [{\citenamefont {Raghu}\ \emph {et~al.}(2010)\citenamefont {Raghu},
  \citenamefont {Kapitulnik},\ and\ \citenamefont
  {Kivelson}}]{raghu2010hidden}%
  \BibitemOpen
  \bibfield  {author} {\bibinfo {author} {\bibfnamefont {S.}~\bibnamefont
  {Raghu}}, \bibinfo {author} {\bibfnamefont {A.}~\bibnamefont {Kapitulnik}},\
  and\ \bibinfo {author} {\bibfnamefont {S.~A.}\ \bibnamefont {Kivelson}},\
  }\bibfield  {title} {\bibinfo {title} {Hidden quasi-one-dimensional
  superconductivity in {${\mathrm{Sr}}_{2}{\mathrm{RuO}}_{4}$}},\ }\href
  {https://doi.org/10.1103/PhysRevLett.105.136401} {\bibfield  {journal}
  {\bibinfo  {journal} {Phys. Rev. Lett.}\ }\textbf {\bibinfo {volume} {105}},\
  \bibinfo {pages} {136401} (\bibinfo {year} {2010})}\BibitemShut {NoStop}%
\bibitem [{\citenamefont {Firmo}\ \emph {et~al.}(2013)\citenamefont {Firmo},
  \citenamefont {Lederer}, \citenamefont {Lupien}, \citenamefont {Mackenzie},
  \citenamefont {Davis},\ and\ \citenamefont {Kivelson}}]{firmo2013Evidence}%
  \BibitemOpen
  \bibfield  {author} {\bibinfo {author} {\bibfnamefont {I.~A.}\ \bibnamefont
  {Firmo}}, \bibinfo {author} {\bibfnamefont {S.}~\bibnamefont {Lederer}},
  \bibinfo {author} {\bibfnamefont {C.}~\bibnamefont {Lupien}}, \bibinfo
  {author} {\bibfnamefont {A.~P.}\ \bibnamefont {Mackenzie}}, \bibinfo {author}
  {\bibfnamefont {J.~C.}\ \bibnamefont {Davis}},\ and\ \bibinfo {author}
  {\bibfnamefont {S.~A.}\ \bibnamefont {Kivelson}},\ }\bibfield  {title}
  {\bibinfo {title} {Evidence from tunneling spectroscopy for a
  quasi-one-dimensional origin of superconductivity in
  {Sr${}_{2}$RuO${}_{4}$}},\ }\href
  {https://doi.org/10.1103/PhysRevB.88.134521} {\bibfield  {journal} {\bibinfo
  {journal} {Phys. Rev. B}\ }\textbf {\bibinfo {volume} {88}},\ \bibinfo
  {pages} {134521} (\bibinfo {year} {2013})}\BibitemShut {NoStop}%
\bibitem [{\citenamefont {Deguchi}\ \emph {et~al.}(2004)\citenamefont
  {Deguchi}, \citenamefont {Mao}, \citenamefont {Yaguchi},\ and\ \citenamefont
  {Maeno}}]{Deguchi2004Gap}%
  \BibitemOpen
  \bibfield  {author} {\bibinfo {author} {\bibfnamefont {K.}~\bibnamefont
  {Deguchi}}, \bibinfo {author} {\bibfnamefont {Z.~Q.}\ \bibnamefont {Mao}},
  \bibinfo {author} {\bibfnamefont {H.}~\bibnamefont {Yaguchi}},\ and\ \bibinfo
  {author} {\bibfnamefont {Y.}~\bibnamefont {Maeno}},\ }\bibfield  {title}
  {\bibinfo {title} {Gap structure of the spin-triplet superconductor
  {${\mathrm{S}\mathrm{r}}_{2}{\mathrm{R}\mathrm{u}\mathrm{O}}_{4}$} determined
  from the field-orientation dependence of the specific heat},\ }\href
  {https://doi.org/10.1103/PhysRevLett.92.047002} {\bibfield  {journal}
  {\bibinfo  {journal} {Phys. Rev. Lett.}\ }\textbf {\bibinfo {volume} {92}},\
  \bibinfo {pages} {047002} (\bibinfo {year} {2004})}\BibitemShut {NoStop}%
\bibitem [{\citenamefont {Bj\"ornsson}\ \emph {et~al.}(2005)\citenamefont
  {Bj\"ornsson}, \citenamefont {Maeno}, \citenamefont {Huber},\ and\
  \citenamefont {Moler}}]{Bjornsson2005Scanning}%
  \BibitemOpen
  \bibfield  {author} {\bibinfo {author} {\bibfnamefont {P.~G.}\ \bibnamefont
  {Bj\"ornsson}}, \bibinfo {author} {\bibfnamefont {Y.}~\bibnamefont {Maeno}},
  \bibinfo {author} {\bibfnamefont {M.~E.}\ \bibnamefont {Huber}},\ and\
  \bibinfo {author} {\bibfnamefont {K.~A.}\ \bibnamefont {Moler}},\ }\bibfield
  {title} {\bibinfo {title} {Scanning magnetic imaging of
  {${\mathrm{Sr}}_{2}\mathrm{Ru}{\mathrm{O}}_{4}$}},\ }\href
  {https://doi.org/10.1103/PhysRevB.72.012504} {\bibfield  {journal} {\bibinfo
  {journal} {Phys. Rev. B}\ }\textbf {\bibinfo {volume} {72}},\ \bibinfo
  {pages} {012504} (\bibinfo {year} {2005})}\BibitemShut {NoStop}%
\bibitem [{\citenamefont {Kirtley}\ \emph {et~al.}(2007)\citenamefont
  {Kirtley}, \citenamefont {Kallin}, \citenamefont {Hicks}, \citenamefont
  {Kim}, \citenamefont {Liu}, \citenamefont {Moler}, \citenamefont {Maeno},\
  and\ \citenamefont {Nelson}}]{Kirtley2007Upper}%
  \BibitemOpen
  \bibfield  {author} {\bibinfo {author} {\bibfnamefont {J.~R.}\ \bibnamefont
  {Kirtley}}, \bibinfo {author} {\bibfnamefont {C.}~\bibnamefont {Kallin}},
  \bibinfo {author} {\bibfnamefont {C.~W.}\ \bibnamefont {Hicks}}, \bibinfo
  {author} {\bibfnamefont {E.-A.}\ \bibnamefont {Kim}}, \bibinfo {author}
  {\bibfnamefont {Y.}~\bibnamefont {Liu}}, \bibinfo {author} {\bibfnamefont
  {K.~A.}\ \bibnamefont {Moler}}, \bibinfo {author} {\bibfnamefont
  {Y.}~\bibnamefont {Maeno}},\ and\ \bibinfo {author} {\bibfnamefont {K.~D.}\
  \bibnamefont {Nelson}},\ }\bibfield  {title} {\bibinfo {title} {Upper limit
  on spontaneous supercurrents in
  {${\mathrm{Sr}}_{2}\mathrm{Ru}{\mathrm{O}}_{4}$}},\ }\href
  {https://doi.org/10.1103/PhysRevB.76.014526} {\bibfield  {journal} {\bibinfo
  {journal} {Phys. Rev. B}\ }\textbf {\bibinfo {volume} {76}},\ \bibinfo
  {pages} {014526} (\bibinfo {year} {2007})}\BibitemShut {NoStop}%
\bibitem [{\citenamefont {Hicks}\ \emph {et~al.}(2010)\citenamefont {Hicks},
  \citenamefont {Kirtley}, \citenamefont {Lippman}, \citenamefont {Koshnick},
  \citenamefont {Huber}, \citenamefont {Maeno}, \citenamefont {Yuhasz},
  \citenamefont {Maple},\ and\ \citenamefont {Moler}}]{Hicks2010Limits}%
  \BibitemOpen
  \bibfield  {author} {\bibinfo {author} {\bibfnamefont {C.~W.}\ \bibnamefont
  {Hicks}}, \bibinfo {author} {\bibfnamefont {J.~R.}\ \bibnamefont {Kirtley}},
  \bibinfo {author} {\bibfnamefont {T.~M.}\ \bibnamefont {Lippman}}, \bibinfo
  {author} {\bibfnamefont {N.~C.}\ \bibnamefont {Koshnick}}, \bibinfo {author}
  {\bibfnamefont {M.~E.}\ \bibnamefont {Huber}}, \bibinfo {author}
  {\bibfnamefont {Y.}~\bibnamefont {Maeno}}, \bibinfo {author} {\bibfnamefont
  {W.~M.}\ \bibnamefont {Yuhasz}}, \bibinfo {author} {\bibfnamefont {M.~B.}\
  \bibnamefont {Maple}},\ and\ \bibinfo {author} {\bibfnamefont {K.~A.}\
  \bibnamefont {Moler}},\ }\bibfield  {title} {\bibinfo {title} {Limits on
  superconductivity-related magnetization in
  {${\text{Sr}}_{2}{\text{RuO}}_{4}$} and {${\text{PrOs}}_{4}{\text{Sb}}_{12}$}
  from scanning squid microscopy},\ }\href
  {https://doi.org/10.1103/PhysRevB.81.214501} {\bibfield  {journal} {\bibinfo
  {journal} {Phys. Rev. B}\ }\textbf {\bibinfo {volume} {81}},\ \bibinfo
  {pages} {214501} (\bibinfo {year} {2010})}\BibitemShut {NoStop}%
\bibitem [{\citenamefont {Goryo}(2008)}]{goryo2008impurity}%
  \BibitemOpen
  \bibfield  {author} {\bibinfo {author} {\bibfnamefont {J.}~\bibnamefont
  {Goryo}},\ }\bibfield  {title} {\bibinfo {title} {Impurity-induced polar kerr
  effect in a chiral $p$-wave superconductor},\ }\href
  {https://doi.org/10.1103/PhysRevB.78.060501} {\bibfield  {journal} {\bibinfo
  {journal} {Phys. Rev. B}\ }\textbf {\bibinfo {volume} {78}},\ \bibinfo
  {pages} {060501} (\bibinfo {year} {2008})}\BibitemShut {NoStop}%
\bibitem [{\citenamefont {Kim}\ \emph {et~al.}(2008)\citenamefont {Kim},
  \citenamefont {Marsiglio},\ and\ \citenamefont {Ting}}]{kim2008hall}%
  \BibitemOpen
  \bibfield  {author} {\bibinfo {author} {\bibfnamefont {W.}~\bibnamefont
  {Kim}}, \bibinfo {author} {\bibfnamefont {F.}~\bibnamefont {Marsiglio}},\
  and\ \bibinfo {author} {\bibfnamefont {C.~S.}\ \bibnamefont {Ting}},\
  }\bibfield  {title} {\bibinfo {title} {Hall conductivity of a spin-triplet
  superconductor},\ }\href {https://doi.org/10.1103/PhysRevLett.100.227003}
  {\bibfield  {journal} {\bibinfo  {journal} {Phys. Rev. Lett.}\ }\textbf
  {\bibinfo {volume} {100}},\ \bibinfo {pages} {227003} (\bibinfo {year}
  {2008})}\BibitemShut {NoStop}%
\bibitem [{\citenamefont {Lutchyn}\ \emph {et~al.}(2009)\citenamefont
  {Lutchyn}, \citenamefont {Nagornykh},\ and\ \citenamefont
  {Yakovenko}}]{lutchyn2009frequency}%
  \BibitemOpen
  \bibfield  {author} {\bibinfo {author} {\bibfnamefont {R.~M.}\ \bibnamefont
  {Lutchyn}}, \bibinfo {author} {\bibfnamefont {P.}~\bibnamefont {Nagornykh}},\
  and\ \bibinfo {author} {\bibfnamefont {V.~M.}\ \bibnamefont {Yakovenko}},\
  }\bibfield  {title} {\bibinfo {title} {Frequency and temperature dependence
  of the anomalous ac hall conductivity in a chiral ${p}_{x}+i{p}_{y}$
  superconductor with impurities},\ }\href
  {https://doi.org/10.1103/PhysRevB.80.104508} {\bibfield  {journal} {\bibinfo
  {journal} {Phys. Rev. B}\ }\textbf {\bibinfo {volume} {80}},\ \bibinfo
  {pages} {104508} (\bibinfo {year} {2009})}\BibitemShut {NoStop}%
\bibitem [{\citenamefont {K\"onig}\ and\ \citenamefont
  {Levchenko}(2017)}]{konig2017kerr}%
  \BibitemOpen
  \bibfield  {author} {\bibinfo {author} {\bibfnamefont {E.~J.}\ \bibnamefont
  {K\"onig}}\ and\ \bibinfo {author} {\bibfnamefont {A.}~\bibnamefont
  {Levchenko}},\ }\bibfield  {title} {\bibinfo {title} {Kerr effect from
  diffractive skew scattering in chiral
  {${p}_{x}\ifmmode\pm\else\textpm\fi{}i{p}_{y}$ }superconductors},\ }\href
  {https://doi.org/10.1103/PhysRevLett.118.027001} {\bibfield  {journal}
  {\bibinfo  {journal} {Phys. Rev. Lett.}\ }\textbf {\bibinfo {volume} {118}},\
  \bibinfo {pages} {027001} (\bibinfo {year} {2017})}\BibitemShut {NoStop}%
\bibitem [{\citenamefont {Taylor}\ and\ \citenamefont
  {Kallin}(2012)}]{taylor2012intrinsic}%
  \BibitemOpen
  \bibfield  {author} {\bibinfo {author} {\bibfnamefont {E.}~\bibnamefont
  {Taylor}}\ and\ \bibinfo {author} {\bibfnamefont {C.}~\bibnamefont
  {Kallin}},\ }\bibfield  {title} {\bibinfo {title} {Intrinsic hall effect in a
  multiband chiral superconductor in the absence of an external magnetic
  field},\ }\href {https://doi.org/10.1103/PhysRevLett.108.157001} {\bibfield
  {journal} {\bibinfo  {journal} {Phys. Rev. Lett.}\ }\textbf {\bibinfo
  {volume} {108}},\ \bibinfo {pages} {157001} (\bibinfo {year}
  {2012})}\BibitemShut {NoStop}%
\bibitem [{\citenamefont {Wysoki\ifmmode~\acute{n}\else \'{n}\fi{}ski}\ \emph
  {et~al.}(2012{\natexlab{a}})\citenamefont {Wysoki\ifmmode~\acute{n}\else
  \'{n}\fi{}ski}, \citenamefont {Annett},\ and\ \citenamefont
  {Gy\"orffy}}]{wysokinski2012intrinsic}%
  \BibitemOpen
  \bibfield  {author} {\bibinfo {author} {\bibfnamefont {K.~I.}\ \bibnamefont
  {Wysoki\ifmmode~\acute{n}\else \'{n}\fi{}ski}}, \bibinfo {author}
  {\bibfnamefont {J.~F.}\ \bibnamefont {Annett}},\ and\ \bibinfo {author}
  {\bibfnamefont {B.~L.}\ \bibnamefont {Gy\"orffy}},\ }\bibfield  {title}
  {\bibinfo {title} {Intrinsic optical dichroism in the chiral superconducting
  state of {${\mathrm{Sr}}_{2}{\mathrm{RuO}}_{4}$}},\ }\href
  {https://doi.org/10.1103/PhysRevLett.108.077004} {\bibfield  {journal}
  {\bibinfo  {journal} {Phys. Rev. Lett.}\ }\textbf {\bibinfo {volume} {108}},\
  \bibinfo {pages} {077004} (\bibinfo {year} {2012}{\natexlab{a}})}\BibitemShut
  {NoStop}%
\bibitem [{\citenamefont {Denys}\ and\ \citenamefont
  {Brydon}(2021)}]{denys2021origin}%
  \BibitemOpen
  \bibfield  {author} {\bibinfo {author} {\bibfnamefont {M.~D.~E.}\
  \bibnamefont {Denys}}\ and\ \bibinfo {author} {\bibfnamefont {P.~M.~R.}\
  \bibnamefont {Brydon}},\ }\bibfield  {title} {\bibinfo {title} {Origin of the
  anomalous hall effect in two-band chiral superconductors},\ }\href
  {https://doi.org/10.1103/PhysRevB.103.094503} {\bibfield  {journal} {\bibinfo
   {journal} {Phys. Rev. B}\ }\textbf {\bibinfo {volume} {103}},\ \bibinfo
  {pages} {094503} (\bibinfo {year} {2021})}\BibitemShut {NoStop}%
\bibitem [{\citenamefont {Wysoki\ifmmode~\acute{n}\else \'{n}\fi{}ski}\ \emph
  {et~al.}(2012{\natexlab{b}})\citenamefont {Wysoki\ifmmode~\acute{n}\else
  \'{n}\fi{}ski}, \citenamefont {Annett},\ and\ \citenamefont
  {Gy\"orffy}}]{ifmmode_2012}%
  \BibitemOpen
  \bibfield  {author} {\bibinfo {author} {\bibfnamefont {K.~I.}\ \bibnamefont
  {Wysoki\ifmmode~\acute{n}\else \'{n}\fi{}ski}}, \bibinfo {author}
  {\bibfnamefont {J.~F.}\ \bibnamefont {Annett}},\ and\ \bibinfo {author}
  {\bibfnamefont {B.~L.}\ \bibnamefont {Gy\"orffy}},\ }\bibfield  {title}
  {\bibinfo {title} {Intrinsic optical dichroism in the chiral superconducting
  state of {${\mathrm{Sr}}_{2}{\mathrm{RuO}}_{4}$}},\ }\href
  {https://doi.org/10.1103/PhysRevLett.108.077004} {\bibfield  {journal}
  {\bibinfo  {journal} {Phys. Rev. Lett.}\ }\textbf {\bibinfo {volume} {108}},\
  \bibinfo {pages} {077004} (\bibinfo {year} {2012}{\natexlab{b}})}\BibitemShut
  {NoStop}%
\bibitem [{\citenamefont {Zhang}\ \emph {et~al.}(2023)\citenamefont {Zhang},
  \citenamefont {Chen}, \citenamefont {Liu}, \citenamefont {Li}, \citenamefont
  {Wang},\ and\ \citenamefont {Huang}}]{Zhang2023Quantum}%
  \BibitemOpen
  \bibfield  {author} {\bibinfo {author} {\bibfnamefont {J.-L.}\ \bibnamefont
  {Zhang}}, \bibinfo {author} {\bibfnamefont {W.}~\bibnamefont {Chen}},
  \bibinfo {author} {\bibfnamefont {H.-T.}\ \bibnamefont {Liu}}, \bibinfo
  {author} {\bibfnamefont {Y.}~\bibnamefont {Li}}, \bibinfo {author}
  {\bibfnamefont {Z.}~\bibnamefont {Wang}},\ and\ \bibinfo {author}
  {\bibfnamefont {W.}~\bibnamefont {Huang}},\ }\href@noop {} {\bibinfo {title}
  {Quantum-geometry-induced anomalous hall effect in non-unitary
  superconductors and application to {Sr$_2$RuO$_4$}}} (\bibinfo {year}
  {2023}),\ \Eprint {https://arxiv.org/abs/2309.14448} {arXiv:2309.14448
  [cond-mat.supr-con]} \BibitemShut {NoStop}%
\bibitem [{\citenamefont {Liu}\ \emph {et~al.}(2023)\citenamefont {Liu},
  \citenamefont {Chen},\ and\ \citenamefont {Huang}}]{Liu2023Impact}%
  \BibitemOpen
  \bibfield  {author} {\bibinfo {author} {\bibfnamefont {H.-T.}\ \bibnamefont
  {Liu}}, \bibinfo {author} {\bibfnamefont {W.}~\bibnamefont {Chen}},\ and\
  \bibinfo {author} {\bibfnamefont {W.}~\bibnamefont {Huang}},\ }\bibfield
  {title} {\bibinfo {title} {Impact of random impurities on the anomalous hall
  effect in chiral superconductors},\ }\href
  {https://doi.org/10.1103/PhysRevB.107.224517} {\bibfield  {journal} {\bibinfo
   {journal} {Phys. Rev. B}\ }\textbf {\bibinfo {volume} {107}},\ \bibinfo
  {pages} {224517} (\bibinfo {year} {2023})}\BibitemShut {NoStop}%
\bibitem [{\citenamefont {Gradhand}\ \emph {et~al.}(2013)\citenamefont
  {Gradhand}, \citenamefont {Wysokinski}, \citenamefont {Annett},\ and\
  \citenamefont {Gy\"orffy}}]{gradhand2013kerr}%
  \BibitemOpen
  \bibfield  {author} {\bibinfo {author} {\bibfnamefont {M.}~\bibnamefont
  {Gradhand}}, \bibinfo {author} {\bibfnamefont {K.~I.}\ \bibnamefont
  {Wysokinski}}, \bibinfo {author} {\bibfnamefont {J.~F.}\ \bibnamefont
  {Annett}},\ and\ \bibinfo {author} {\bibfnamefont {B.~L.}\ \bibnamefont
  {Gy\"orffy}},\ }\bibfield  {title} {\bibinfo {title} {Kerr rotation in the
  unconventional superconductor {Sr${}_{2}$RuO${}_{4}$}},\ }\href
  {https://doi.org/10.1103/PhysRevB.88.094504} {\bibfield  {journal} {\bibinfo
  {journal} {Phys. Rev. B}\ }\textbf {\bibinfo {volume} {88}},\ \bibinfo
  {pages} {094504} (\bibinfo {year} {2013})}\BibitemShut {NoStop}%
\bibitem [{\citenamefont {Ramires}\ and\ \citenamefont
  {Sigrist}(2016)}]{Ramires2016Identifying}%
  \BibitemOpen
  \bibfield  {author} {\bibinfo {author} {\bibfnamefont {A.}~\bibnamefont
  {Ramires}}\ and\ \bibinfo {author} {\bibfnamefont {M.}~\bibnamefont
  {Sigrist}},\ }\bibfield  {title} {\bibinfo {title} {Identifying detrimental
  effects for multiorbital superconductivity: Application to
  {${\mathrm{Sr}}_{2}{\mathrm{RuO}}_{4}$}},\ }\href
  {https://doi.org/10.1103/PhysRevB.94.104501} {\bibfield  {journal} {\bibinfo
  {journal} {Phys. Rev. B}\ }\textbf {\bibinfo {volume} {94}},\ \bibinfo
  {pages} {104501} (\bibinfo {year} {2016})}\BibitemShut {NoStop}%
\bibitem [{\citenamefont {Ramires}\ and\ \citenamefont
  {Sigrist}(2017)}]{Ramires2017Notes}%
  \BibitemOpen
  \bibfield  {author} {\bibinfo {author} {\bibfnamefont {A.}~\bibnamefont
  {Ramires}}\ and\ \bibinfo {author} {\bibfnamefont {M.}~\bibnamefont
  {Sigrist}},\ }\bibfield  {title} {\bibinfo {title} {A note on the upper
  critical field of {${\mathrm{Sr}}_{2}{\mathrm{RuO}}_{4}$} under strain},\
  }\href {https://doi.org/10.1088/1742-6596/807/5/052011} {\ \textbf {\bibinfo
  {volume} {807}},\ \bibinfo {pages} {052011} (\bibinfo {year}
  {2017})}\BibitemShut {NoStop}%
\bibitem [{\citenamefont {Wang}\ \emph {et~al.}(2020)\citenamefont {Wang},
  \citenamefont {Wang},\ and\ \citenamefont {Kallin}}]{Kallin2020Spin}%
  \BibitemOpen
  \bibfield  {author} {\bibinfo {author} {\bibfnamefont {Z.}~\bibnamefont
  {Wang}}, \bibinfo {author} {\bibfnamefont {X.}~\bibnamefont {Wang}},\ and\
  \bibinfo {author} {\bibfnamefont {C.}~\bibnamefont {Kallin}},\ }\bibfield
  {title} {\bibinfo {title} {Spin-orbit coupling and spin-triplet pairing
  symmetry in {${\mathrm{Sr}}_{2}{\mathrm{RuO}}_{4}$ }},\ }\href
  {https://doi.org/10.1103/PhysRevB.101.064507} {\bibfield  {journal} {\bibinfo
   {journal} {Phys. Rev. B}\ }\textbf {\bibinfo {volume} {101}},\ \bibinfo
  {pages} {064507} (\bibinfo {year} {2020})}\BibitemShut {NoStop}%
\bibitem [{\citenamefont {Akbari}\ and\ \citenamefont
  {Thalmeier}(2013)}]{akbari2013}%
  \BibitemOpen
  \bibfield  {author} {\bibinfo {author} {\bibfnamefont {A.}~\bibnamefont
  {Akbari}}\ and\ \bibinfo {author} {\bibfnamefont {P.}~\bibnamefont
  {Thalmeier}},\ }\bibfield  {title} {\bibinfo {title} {Multiorbital and
  hybridization effects in the quasiparticle interference of the triplet
  superconductor {Sr${}_{2}$RuO${}_{4}$}},\ }\href
  {https://doi.org/10.1103/PhysRevB.88.134519} {\bibfield  {journal} {\bibinfo
  {journal} {Phys. Rev. B}\ }\textbf {\bibinfo {volume} {88}},\ \bibinfo
  {pages} {134519} (\bibinfo {year} {2013})}\BibitemShut {NoStop}%
\bibitem [{\citenamefont {Taylor}\ and\ \citenamefont
  {Kallin}(2013)}]{taylor2013anomalous}%
  \BibitemOpen
  \bibfield  {author} {\bibinfo {author} {\bibfnamefont {E.}~\bibnamefont
  {Taylor}}\ and\ \bibinfo {author} {\bibfnamefont {C.}~\bibnamefont
  {Kallin}},\ }\bibfield  {title} {\bibinfo {title} {Anomalous hall
  conductivity of clean {${\mathrm{Sr}}_{2}{\mathrm{RuO}}_{4}$} at finite
  temperatures},\ }\href {https://doi.org/10.1088/1742-6596/449/1/012036}
  {\bibfield  {journal} {\bibinfo  {journal} {Journal of Physics: Conference
  Series}\ }\textbf {\bibinfo {volume} {449}},\ \bibinfo {pages} {012036}
  (\bibinfo {year} {2013})}\BibitemShut {NoStop}%
\bibitem [{\citenamefont {Scaffidi}(2023)}]{scaffidi2020degeneracy}%
  \BibitemOpen
  \bibfield  {author} {\bibinfo {author} {\bibfnamefont {T.}~\bibnamefont
  {Scaffidi}},\ }\bibfield  {title} {\bibinfo {title} {Degeneracy between even-
  and odd-parity superconductivity in the quasi-1d hubbard model and
  implications for {${\mathrm{Sr}}_{2}{\mathrm{RuO}}_{4}$}},\ }\href@noop {} {\
   (\bibinfo {year} {2023})},\ \Eprint {https://arxiv.org/abs/2007.13769}
  {arXiv:2007.13769 [cond-mat.supr-con]} \BibitemShut {NoStop}%
\bibitem [{\citenamefont {Palle}\ \emph {et~al.}(2023)\citenamefont {Palle},
  \citenamefont {Hicks}, \citenamefont {Valent\'{\i}}, \citenamefont {Hu},
  \citenamefont {Li}, \citenamefont {Rost}, \citenamefont {Nicklas},
  \citenamefont {Mackenzie},\ and\ \citenamefont
  {Schmalian}}]{Palle2023Constraints}%
  \BibitemOpen
  \bibfield  {author} {\bibinfo {author} {\bibfnamefont {G.}~\bibnamefont
  {Palle}}, \bibinfo {author} {\bibfnamefont {C.}~\bibnamefont {Hicks}},
  \bibinfo {author} {\bibfnamefont {R.}~\bibnamefont {Valent\'{\i}}}, \bibinfo
  {author} {\bibfnamefont {Z.}~\bibnamefont {Hu}}, \bibinfo {author}
  {\bibfnamefont {Y.-S.}\ \bibnamefont {Li}}, \bibinfo {author} {\bibfnamefont
  {A.}~\bibnamefont {Rost}}, \bibinfo {author} {\bibfnamefont {M.}~\bibnamefont
  {Nicklas}}, \bibinfo {author} {\bibfnamefont {A.~P.}\ \bibnamefont
  {Mackenzie}},\ and\ \bibinfo {author} {\bibfnamefont {J.}~\bibnamefont
  {Schmalian}},\ }\bibfield  {title} {\bibinfo {title} {Constraints on the
  superconducting state of {${\text{Sr}}_{2}{\text{RuO}}_{4}$} from
  elastocaloric measurements},\ }\href
  {https://doi.org/10.1103/PhysRevB.108.094516} {\bibfield  {journal} {\bibinfo
   {journal} {Phys. Rev. B}\ }\textbf {\bibinfo {volume} {108}},\ \bibinfo
  {pages} {094516} (\bibinfo {year} {2023})}\BibitemShut {NoStop}%
\bibitem [{\citenamefont {Wang}\ \emph {et~al.}(2022)\citenamefont {Wang},
  \citenamefont {Wang},\ and\ \citenamefont {Kallin}}]{Wang2022Higher}%
  \BibitemOpen
  \bibfield  {author} {\bibinfo {author} {\bibfnamefont {X.}~\bibnamefont
  {Wang}}, \bibinfo {author} {\bibfnamefont {Z.}~\bibnamefont {Wang}},\ and\
  \bibinfo {author} {\bibfnamefont {C.}~\bibnamefont {Kallin}},\ }\bibfield
  {title} {\bibinfo {title} {Higher angular momentum pairing states in
  {${\mathrm{Sr}}_{2}{\mathrm{RuO}}_{4}$} in the presence of longer-range
  interactions},\ }\href {https://doi.org/10.1103/PhysRevB.106.134512}
  {\bibfield  {journal} {\bibinfo  {journal} {Phys. Rev. B}\ }\textbf {\bibinfo
  {volume} {106}},\ \bibinfo {pages} {134512} (\bibinfo {year}
  {2022})}\BibitemShut {NoStop}%
\bibitem [{\citenamefont {Yanase}\ and\ \citenamefont
  {Ogata}(2003)}]{yanase2003microscopic}%
  \BibitemOpen
  \bibfield  {author} {\bibinfo {author} {\bibfnamefont {Y.}~\bibnamefont
  {Yanase}}\ and\ \bibinfo {author} {\bibfnamefont {M.}~\bibnamefont {Ogata}},\
  }\bibfield  {title} {\bibinfo {title} {Microscopic identification of the
  d-vector in triplet superconductor {{${\mathrm{Sr}}_{2}{\mathrm{RuO}}_{4}$}
  }},\ }\href {https://doi.org/10.1143/JPSJ.72.673} {\bibfield  {journal}
  {\bibinfo  {journal} {Journal of the Physical Society of Japan}\ }\textbf
  {\bibinfo {volume} {72}},\ \bibinfo {pages} {673} (\bibinfo {year}
  {2003})}\BibitemShut {NoStop}%
\bibitem [{\citenamefont {Haverkort}\ \emph {et~al.}(2008)\citenamefont
  {Haverkort}, \citenamefont {Elfimov}, \citenamefont {Tjeng}, \citenamefont
  {Sawatzky},\ and\ \citenamefont {Damascelli}}]{haverkort2008strong}%
  \BibitemOpen
  \bibfield  {author} {\bibinfo {author} {\bibfnamefont {M.~W.}\ \bibnamefont
  {Haverkort}}, \bibinfo {author} {\bibfnamefont {I.~S.}\ \bibnamefont
  {Elfimov}}, \bibinfo {author} {\bibfnamefont {L.~H.}\ \bibnamefont {Tjeng}},
  \bibinfo {author} {\bibfnamefont {G.~A.}\ \bibnamefont {Sawatzky}},\ and\
  \bibinfo {author} {\bibfnamefont {A.}~\bibnamefont {Damascelli}},\ }\bibfield
   {title} {\bibinfo {title} {Strong spin-orbit coupling effects on the fermi
  surface of {${\mathrm{Sr}}_{2}{\mathrm{RuO}}_{4}$ and
  ${\mathrm{Sr}}_{2}{\mathrm{RhO}}_{4}$}},\ }\href
  {https://doi.org/10.1103/PhysRevLett.101.026406} {\bibfield  {journal}
  {\bibinfo  {journal} {Phys. Rev. Lett.}\ }\textbf {\bibinfo {volume} {101}},\
  \bibinfo {pages} {026406} (\bibinfo {year} {2008})}\BibitemShut {NoStop}%
\bibitem [{\citenamefont {Katsufuji}\ \emph {et~al.}(1996)\citenamefont
  {Katsufuji}, \citenamefont {Kasai},\ and\ \citenamefont
  {Tokura}}]{katsufuji1996plane}%
  \BibitemOpen
  \bibfield  {author} {\bibinfo {author} {\bibfnamefont {T.}~\bibnamefont
  {Katsufuji}}, \bibinfo {author} {\bibfnamefont {M.}~\bibnamefont {Kasai}},\
  and\ \bibinfo {author} {\bibfnamefont {Y.}~\bibnamefont {Tokura}},\
  }\bibfield  {title} {\bibinfo {title} {In-plane and out-of-plane optical
  spectra of {${\mathrm{Sr}}_{2}{\mathrm{RuO}}_{4}$}},\ }\href
  {https://doi.org/10.1103/PhysRevLett.76.126} {\bibfield  {journal} {\bibinfo
  {journal} {Phys. Rev. Lett.}\ }\textbf {\bibinfo {volume} {76}},\ \bibinfo
  {pages} {126} (\bibinfo {year} {1996})}\BibitemShut {NoStop}%
\end{thebibliography}%
%
%
\end{document}